\documentclass[amsmath,amssymb,aps,pra,10pt,twocolumn,groupedaddress,nofootinbib, floatfix, superscriptaddress]{revtex4-1}

\usepackage{graphicx}
\usepackage{bm,bbm}
\usepackage{hyperref}
\usepackage[margin=2cm]{geometry}
\usepackage{xcolor}
\usepackage{qcircuit}
\usepackage{flafter}

\newcommand{\ket}[1]{| #1 \rangle} 
\newcommand{\bra}[1]{\langle #1 |} 
\newcommand{\braket}[2]{\langle #1 | #2 \rangle} 
\newcommand{\ketbra}[2]{|#1\rangle\langle#2|}
\renewcommand{\a}{\hat{a}}
\newcommand{\ad}{\a^\dagger}
\newcommand{\x}{\hat{x}}
\newcommand{\p}{\hat{p}}
\newcommand{\bx}{\mathbf{x}}
\newcommand{\bp}{\mathbf{p}}

\newcommand{\beq}{\begin{equation}}
\newcommand{\eeq}{\end{equation}}

\newcommand{\bad}{\bm{\hat a}^\dagger}

\begin{document}

\title{Continuous-variable quantum neural networks}

\author{Nathan Killoran}
\affiliation{Xanadu, 372 Richmond St W, Toronto, M5V 1X6, Canada}
\author{Thomas R. Bromley}
\affiliation{Xanadu, 372 Richmond St W, Toronto, M5V 1X6, Canada}
\author{Juan Miguel Arrazola}
\affiliation{Xanadu, 372 Richmond St W, Toronto, M5V 1X6, Canada}
\author{Maria Schuld}
\affiliation{Xanadu, 372 Richmond St W, Toronto, M5V 1X6, Canada}
\author{Nicol\'{a}s Quesada}
\affiliation{Xanadu, 372 Richmond St W, Toronto, M5V 1X6, Canada}
\author{Seth Lloyd}
\affiliation{Massachusetts  Institute  of  Technology,  Department  of  Mechanical  Engineering,
77  Massachusetts  Avenue,  Cambridge,  Massachusetts  02139,  USA}

\date{\today}

\begin{abstract}
We introduce a general method for building neural networks on quantum computers.
The quantum neural network is a variational quantum circuit built in the continuous-variable (CV) architecture, which encodes quantum information in continuous degrees of freedom such as the amplitudes of the electromagnetic field. 
This circuit contains a layered structure of continuously parameterized gates which is universal for CV quantum computation. 
Affine transformations and nonlinear activation functions, two key elements in neural networks, are enacted in the quantum network using Gaussian and non-Gaussian gates, respectively. 
The non-Gaussian gates provide both the nonlinearity and the universality of the model. 
Due to the structure of the CV model, the CV quantum neural network can encode highly nonlinear transformations while remaining completely unitary. 
We show how a classical network can be embedded into the quantum formalism and propose quantum versions of various specialized model such as convolutional, recurrent, and residual networks.
Finally, we present numerous modeling experiments built with the Strawberry Fields software library.
These experiments, including a classifier for fraud detection, a network which generates Tetris images, and a hybrid classical-quantum autoencoder, demonstrate the capability and adaptability of CV quantum neural networks. 
\end{abstract}

\maketitle

\section{Introduction}
\label{sec:intro}
After many years of scientific development, quantum computers are now beginning to move out of the lab and into the mainstream. Over those years of research, many powerful algorithms and applications for quantum hardware have been established. In particular, the potential for quantum computers to enhance machine learning is truly exciting \cite{biamonte2017quantum, wittek2014quantum, schuld2018book}. Sufficiently powerful quantum computers can in principle provide computational speedups for key machine learning algorithms and subroutines such as data fitting \cite{wiebe2012quantum}, principal component analysis \cite{lloyd2014quantum}, Bayesian inference \cite{low2014quantum,wiebe2015can}, Monte Carlo methods \cite{montanaro2015quantum}, support vector machines \cite{rebentrost2014quantum, schuld2018quantum}, Boltzmann machines \cite{amin2018quantum, kieferova2016tomography}, and recommendation systems \cite{kerenidis2016quantum}. 

On the classical computing side, there has recently been a renaissance in machine learning techniques based on neural networks, forming the new field of deep learning \cite{lecun2015deep, schmidhuber2015deep, goodfellow2016deep}. This breakthrough is being fueled by a number of technical factors, including new software libraries \cite{bergstra2010theano, jia2014caffe, maclaurin2015autograd, abadi2016tensorflow, paszke2017automatic} and powerful special-purpose computational hardware \cite{chetlur2014cudnn, jouppi2017datacenter}. Rather than the conventional bit registers found in digital computing, the fundamental computational units in deep learning are continuous vectors and tensors which are transformed in high-dimensional spaces. At the moment, these continuous computations are still approximated using conventional digital computers. However, new specialized computational hardware is currently being engineered which is fundamentally analog in nature \cite{mead1990neuromorphic, poon2011neuromorphic, appeltant2011information, tait2014broadcast, monroe2014neuromorphic, tait2014photonic, vandoorne2014experimental, shen2017deep}. 

Quantum computation is a paradigm that furthermore includes nonclassical effects such as superposition, interference, and entanglement, giving it potential advantages over classical computing models. Together, these ingredients make quantum computers an intriguing platform for exploring new types of neural networks, in particular hybrid classical-quantum schemes \cite{romero2017quantum, wan2017quantum, verdon2017quantum, farhi2018classification, mitarai2018quantum, schuld2018circuit, grant2018hierarchical, chen2018universal}. Yet the familiar qubit-based quantum computer has the drawback that it is not wholly continuous, since the measurement outputs of qubit-based circuits are generally discrete. Rather, it can be thought of as a type of \emph{digital} quantum hardware \cite{adesso2014continuous}, only partially suited to continuous-valued problems \cite{perdomo2017opportunities,benedetti2018quantum}. 

The quantum computing architecture which is most naturally continuous is the \emph{continuous-variable (CV)} model. Intuitively, the CV model leverages the wave-like properties of nature. Quantum information is encoded not in qubits, but in the quantum states of fields, such as the electromagnetic field, making it ideally suited to photonic hardware. 
The standard observables in the CV picture, e.g., position $\hat{x}$ or momentum $\hat{p}$, have continuous outcomes. Importantly, qubit computations can be embedded into the quantum field picture \cite{gottesman2001encoding,knill2001scheme}, so there is no loss in computational power by taking the CV approach.
Recently, the first steps towards using the CV model for machine learning have begun to be explored, showing how several basic machine learning primitives 
can be built in the CV setting \cite{lau2017quantum, das2018continuous}. As well, a kernel-based classifier using a CV quantum circuit was trained in \cite{schuld2018quantum}. Beyond these early forays, the CV model remains largely unexplored territory as a setting for machine learning. 

In this work, we show that the CV model gives a native architecture for building neural network models on quantum computers. We propose a variational quantum circuit which straightforwardly extends the notion of a fully connected layer structure from classical neural networks to the quantum realm. This quantum circuit contains a continuously parameterized set of operations which are universal for CV quantum computation. By stacking multiple building blocks of this type, we can create multilayer quantum networks which are increasingly expressive. Since the network is made from a universal set of gates, this architecture can also provide a quantum advantage: for certain problems, a classical neural network would require exponentially many resources to approximate the quantum network. Furthermore, we show how to embed classical neural networks into a CV quantum network by restricting to the special case where the gates and parameters of the network do not create any superposition or entanglement. 

This paper is organized as follows. In Sec. \ref{sec:overview}, we review the key concepts from deep learning and from quantum computing which set up the remainder of the paper. We then introduce our basic continuous-variable quantum neural network model in Sec. \ref{sec:qnn} and explore it in detail theoretically. In Sec. \ref{sec:numerics}, we validate and showcase the CV quantum neural network architecture through several machine learning modeling experiments. We conclude with some final thoughts in Sec. \ref{sec:conclusion}.

\section{Overview}
\label{sec:overview}
In this section, we give a high-level synopsis of both deep learning and the CV model. To make this work more accessible to practitioners from diverse backgrounds, we will defer the more technical points to later sections. Both deep learning and CV quantum computation are rich fields; further details can be found in various review papers and textbooks \cite{lecun2015deep, goodfellow2016deep, ferraro2005gaussian, weedbrook2012gaussian, adesso2014continuous, serafini2017quantum}.

\subsection{Neural networks and deep learning}

The fundamental construct in deep learning is the \emph{feedforward neural network} (also known as the \emph{multilayer perceptron}) \cite{goodfellow2016deep}. Over time, this key element has been augmented with additional structure -- such as convolutional feature maps \cite{lecun1989backpropagation}, recurrent connections \cite{rumelhart1986learning}, attention mechanisms \cite{ba2014multiple}, or external memory \cite{graves2014neural} -- for more specialized or advanced use cases. Yet the basic recipe remains largely the same: a multilayer structure, where each layer consists of a linear transformation followed by a nonlinear `activation' function. Mathematically, for an input vector $\bx\in\mathbb{R}^n$, a single layer $\mathcal{L}$ performs the transformation
\begin{equation}
  \mathcal{L}(\bx) = \varphi(W\bx + \mathbf{b}),
  \label{eq:single_layer}
\end{equation}
where $W\in\mathbb{R}^{m\times n}$ is a matrix, $\mathbf{b}\in\mathbb{R}^m$ is a vector, and $\varphi$ is the nonlinear function. The objects $W$ and $\mathbf{b}$ -- called the \emph{weight matrix} and the \emph{bias vector}, respectively -- are made up of free parameters $\theta_W$ and $\theta_b$. Typically, the activation function $\varphi$ contains no free parameters and acts element-wise on its inputs. 

The `deep' in deep learning comes from stacking multiple layers of this type together, so that the output of one layer is used as an input for the next. In general, each layer $\mathcal{L}_i$ will have its own independent weight and bias parameters. Summarizing all model parameters by the parameter set $\bm{\theta}$, an $N$-layer neural network model is given by 
\begin{equation}
  \mathbf{y} = f_{\bm{\theta}}(\bx) = \mathcal{L}_{N} \circ \cdots \circ \mathcal{L}_{1}(\bx),
  \label{eq:base_model}
\end{equation}
and maps an input $\bx$ to a final output $\mathbf{y}$.

Building machine learning models with multilayer neural networks is well-motivated because of various universality theorems \cite{hornik1989multilayer, cybenko1989approximation, leshno1993multilayer}. These theorems guarantee that, provided enough free parameters, feedforward neural networks can approximate any continuous function on a closed and bounded subset of $\mathbb{R}^n$ to an arbitrary degree of accuracy. While the original theorems showed that two layers were sufficient for universal function approximation, deeper networks can be more powerful and more efficient than shallower networks with the same number of parameters \cite{maass1994comparison, montufar2014universal, lin2017does}. 

The universality theorems prove the power of the neural network model for approximating functions, but those theorems do not say anything about how to actually find this approximation. Typically, the function to be fitted is not explicitly known, but rather its input-output relation is to be inferred from data. How can we adjust the network parameters so that it fits the given data?  For this task, the workhorse is the stochastic gradient descent algorithm \cite{bottou1998online}, which fits a neural network model to data by estimating derivatives of the model's parameters -- the weights and biases -- and using gradient descent to minimize some relevant objective function. Combined with a sufficiently large dataset, neural networks trained via stochastic gradient descent have shown remarkable performance for a variety of tasks across many application areas \cite{lecun2015deep, goodfellow2016deep}. 

\subsection{Quantum computing and the CV model}

The quantum analogue of the classical bit is the qubit. The quantum states of a many-qubit system are normalized vectors in a complex Hilbert space. 
Various attempts have been made over the years to encode neural networks and neural-network-like structures into qubit systems, with varying degrees of success \cite{schuld2014quest}. 
One can roughly distinguish two strategies. There are approaches that encode inputs into the amplitude vector of a multiqubit state and interpret unitary transformations as neural network layers. These models require indirect techniques to introduce the crucial nonlinearity of the activation function, which often lead to a nonnegligible probability for the algorithm to fail \cite{torrontegui18, cao17, schuld18cc}. Other approaches, which encode each input bit into a separate qubit \cite{farhi18, wan17}, have an overhead stemming from the need to binarize the continuous values. Furthermore, the typical neural network structure of matrix multiplication and nonlinear activations becomes cumbersome to translate into a quantum algorithm, and the advantages of doing so are not always apparent.
Due to these constraints, qubit architectures are arguably not the most flexible quantum frameworks for encoding neural networks, which have continuous real-valued inputs and outputs.

Fortunately, qubits are not the sole medium available for quantum information processing. An alternate quantum computing architecture, the CV model \cite{lloyd1999quantum}, is a much better fit with the continuous picture of computation underlying neural networks. The CV formalism has a long history, and can be physically realized using optical systems \cite{andersen2015hybrid,yoshikawa2016invited}, in the microwave regime \cite{moon2005theory,peropadre2016proposal,girvin2017schrodinger}, and using ion traps \cite{shen2014scalable,meekhof1996generation, monroe1996schrodinger}. 
In the CV model, information is carried in the quantum states of bosonic modes, often called \emph{qumodes}, which form the `wires' of a quantum circuit. 
Continuous-variable quantum information can be encoded using two related pictures: the \emph{wavefunction representation} \cite{schrodinger1926quantisierung, schrodinger1926undulatory} and the \emph{phase space formulation} of quantum mechanics \cite{weyl1927quantenmechanik,wigner1932quantum,groenewold1946principles, moyal1949quantum}. In the former, we specify a single continuous variable, say $x$, and represent the state of the qumode through a complex-valued function of this variable called the wavefunction $\psi(x)$. Concretely, we can interpret $x$ as a position coordinate, and $|\psi(x)|^2$ as the probability density of a particle being located at $x$. From elementary quantum theory, we can also use a wavefunction based on a conjugate momentum variable, $\phi(p)$. Instead of position and momentum, $x$ and $p$ can equivalently be pictured as the real and imaginary parts of a quantum field, such as light. 

In the phase space picture, we treat the conjugate variables $x$ and $p$ on equal footing, giving a connection to classical Hamiltonian mechanics. Thus, the state of a single qumode is encoded with two real-valued variables $(x,p)\in\mathbb{R}^2$. For $N$-qumodes, the phase space employs $2N$ real variables $(\bx,\bp)\in \mathbb{R}^{2N}$. Qumode states are represented as real-valued functions $F(\bx,\bp)$ in phase space called \emph{quasiprobability distributions}. `Quasi' refers to the fact that these functions share some, but not all, properties with classical probability distributions. Specifically, quasiprobability functions can be negative. While normalization forces qubit systems to have a unitary geometry, normalization gives a much looser constraint in the CV picture, namely that the function $F(\bx,\bp)$ has unit integral over the phase space. 
Qumode states also have a representation as vectors or density matrices in the countably infinite Hilbert space spanned by the \emph{Fock states} $\{|n\rangle\}_{n=0}^\infty$, which are the eigenstates of the photon number operator $\hat{n}$. These basis states represent the particle-like nature of qumode systems, with $n$ denoting the number of particles. This is analogous to how square-integrable functions can be expanded using a countable basis set like sines or cosines. 

The phase space and Hilbert space formulations give equivalent predictions. Thus, CV quantum systems can be explored from both a wave-like and a particle-like perspective. We will mainly concentrate on the former.

\subsubsection*{Gaussian operations}

There is a key distinction in the CV model between the quantum gates which are \emph{Gaussian} and those which are not. In many ways, the Gaussian gates are the ``easy'' operations for a CV quantum computer. The simplest single-mode Gaussian gates are \emph{rotation} $R(\phi)$, \emph{displacement} $D(\alpha)$, and \emph{squeezing} $S(r)$. The basic two-mode Gaussian gate is the (phaseless) \emph{beamsplitter} $BS(\theta)$, which can be understood as a rotation between two qumodes. More explicitly, these Gaussian gates produce the following transformations on phase space:
\begin{align}
 & R(\phi): &
  \begin{bmatrix}
    x \\ 
    p
  \end{bmatrix}
  \mapsto &
  \begin{bmatrix}
    \cos\phi & \sin\phi \\ 
    -\sin\phi & \cos\phi
  \end{bmatrix}
  \begin{bmatrix}
    x \\ 
    p
  \end{bmatrix}, \label{eq:rotation}
 \\
 & D(\alpha): &
  \begin{bmatrix}
    x \\ 
    p
  \end{bmatrix}
  \mapsto &
  \begin{bmatrix}
    x + \mathrm{Re}(\alpha)\\ 
    p + \mathrm{Im}(\alpha)
  \end{bmatrix}, \label{eq:displacement}
 \\
 & S(r): &
  \begin{bmatrix}
    x \\ 
    p
  \end{bmatrix}
  \mapsto &
  \begin{bmatrix}
    e^{-r} & 0 \\ 
    0 & e^{r}
  \end{bmatrix}
  \begin{bmatrix}
    x \\ 
    p
  \end{bmatrix}, \label{eq:squeezing}
 \\
 & BS(\theta): & 
  \begin{bmatrix}
    x_1 \\ 
    x_2 \\
    p_1 \\
    p_2 \\
  \end{bmatrix}
  \mapsto &
  \begin{bmatrix}
    \cos\theta & -\sin\theta & 0 & 0\\ 
    \sin\theta & \cos\theta & 0 & 0\\
    0 & 0 & \cos\theta & -\sin\theta \\
    0 & 0 & \sin\theta & \cos\theta
  \end{bmatrix}
  \begin{bmatrix}
    x_1 \\ 
    x_2 \\
    p_1 \\
    p_2 \\
  \end{bmatrix}. \label{eq:beamsplitter}
\end{align}
The ranges for the parameter values are $\phi,\theta\in[0,2\pi]$, $\alpha\in\mathbb{C}\cong\mathbb{R}^2$, and $r\in\mathbb{R}$.

Notice that most of these Gaussian operations have names suggestive of a linear character. Indeed, there is a natural correspondence between Gaussian operations and affine transformations on phase space. For a system of $N$ modes, the most general Gaussian transformation has the effect
\begin{equation}
\label{eq:gen_gaussian_transf}
  \begin{bmatrix}
    \bx \\ 
    \bp
  \end{bmatrix}
  \mapsto
  M
  \begin{bmatrix}
    \bx \\ 
    \bp
  \end{bmatrix}
  + 
  \begin{bmatrix}
    \bm{\alpha}_r \\ 
    \bm{\alpha}_i
  \end{bmatrix},
\end{equation}
where $M$ is a real-valued \emph{symplectic matrix} and $\bm{\alpha}\in\mathbb{C}^N\cong\mathbb{R}^{2N}$ is a complex vector with real/imaginary parts $\bm{\alpha}_r/\bm{\alpha}_i$. This native affine structure will be our key for building quantum neural networks.

A matrix $M$ is symplectic if it satisfies the relation $M^T\Omega M = \Omega$ where
\begin{equation}
 \label{eq:symplectic_form}
 \Omega = 
  \begin{bmatrix}
    0 & \mathbbm{1} \\
    -\mathbbm{1} & 0
  \end{bmatrix}
\end{equation}
is the $2N\times 2N$ \emph{symplectic form}. 
A generic symplectic matrix $M$ can be split into a type of singular-value decomposition -- known as the \emph{Euler} or \emph{Bloch-Messiah} decomposition \cite{weedbrook2012gaussian, serafini2017quantum} -- of the form
\begin{equation}
 \label{eq:bloch_messiah}
 M = K_2 
 \begin{bmatrix}
   \Sigma & 0 \\
   0 & \Sigma^{-1}
 \end{bmatrix}
 K_1,
\end{equation}
where $\Sigma=\mathrm{diag}(c_1,\dots,c_N)$ with $c_i>0$, and $K_1$ and $K_2$ are real-valued matrices which are symplectic and orthogonal. A matrix $K$ with these two properties must have the form
\begin{equation}
 \label{eq:symplectic_orthogonal_block}
 K =  
 \begin{bmatrix}
   C & D \\
  -D & C
 \end{bmatrix},
\end{equation}
with 
\begin{align}
 CD^T - DC^T = 0 \label{eq:symplectic_orthogonal_eqns1} \\
 CC^T + DD^T = \mathbbm{1}. \label{eq:symplectic_orthogonal_eqns2}
\end{align}
We will also need later the fact that if $C$ is an arbitrary orthogonal matrix, then $C\oplus C$ is both orthogonal and symplectic.
Importantly, the intersection of the symplectic and orthogonal groups on $2N$ dimensions is isomorphic to the unitary group on $N$ dimensions. This isomorphism allows us to perform the transformations $K_i$ via the unitary action of passive linear optical interferometers.

Every Gaussian transformation on $N$ modes (Eq. (\ref{eq:gen_gaussian_transf})) can be decomposed into a CV circuit containing only the basic gates mentioned above.
Looking back to Eqs. (\ref{eq:rotation})-(\ref{eq:beamsplitter}), we can recognize that interferometers made up of $R$ and $BS$ gates are sufficient to generate the orthogonal transformations $K_1$, $K_2$, while $S$ gates are sufficient to give the scaling transformation $\Sigma\oplus\Sigma^{-1}$. Finally, displacement gates complete the full affine transformation. Alternatively, we could have defined the Gaussian transformations as those quantum circuits which contain only the gates given above. The Gaussian transformations are so-named because they map the set of Gaussian distributions in phase space to itself. 

\begin{table}
  \begin{tabular}{ l l }
    Classical & CV quantum computing \\
    \hline
    feedforward neural network & CV variational circuit \\
    weight matrix $W$ & symplectic matrix $M$ \\
    bias vector $\bm{b}$ & displacement vector $\bm{\alpha}$ \\
    affine transformations & Gaussian gates \\
    nonlinear function & non-Gaussian gate \\
    weight/bias parameters & gate parameters \\
    variable $x$ & operator $\x$ \\
    derivative $\frac{\partial}{\partial x}$ & conjugate operator $\p$ \\
    no classical analogue & superposition \\
    no classical analogue & entanglement \\
  \end{tabular}
  \caption{Conceptual correspondences between classical neural networks and CV quantum computing. Some concepts from the quantum side have no classical analogue.}
  \label{tab:correspond}
\end{table}

\subsubsection*{Universality in the CV model}

Similar to neural networks, quantum computing comes with its own inherent notions of `universality.'
To define universality in the CV model, we need to first introduce operator versions of the phase space variables, namely $\x$ and $\p$. The $\x$ operator has a spectrum consisting of the entire real line:
\begin{equation}
 \x = \int_{-\infty}^\infty x \ketbra{x}{x} dx,
\end{equation}
where the vectors $\ket{x}$ are orthogonal, $\braket{x}{x'}=\delta(x-x')$. This operator is not trace-class, and the vectors $\ket{x}$ are not normalizable. In the phase space representation, the eigenstates $\ket{x'}$ correspond to ellipses centered at $x=x'$ which are infinitely squeezed, i.e., infinitesimal along the $x$-axis and correspondingly infinite in extent on the $p$-axis.
The conjugate operator $\p$ has a similar structure:
\begin{equation}
 \p = \int_{-\infty}^\infty p \ketbra{p}{p} dp,
\end{equation}
where $\braket{p}{p'}=\delta(p-p')$ and $\braket{p}{x} \sim e^{-ipx}$.
Each qumode of a CV quantum computer is associated with a pair of operators $(\x_i, \p_i)$. For multiple modes, we combine the associated operators together into vectors $(\mathbf{\x}, \mathbf{\p})$.

These operators have the commutator $[\x_j,\p_k]=i\Omega_{jk}$, which leads to the famous uncertainty relation for simultaneous measurements of $\x$ and $\p$.
Connecting to Eq. (\ref{eq:rotation}), we can associate $\p$ with a rotation of the operator $\x$; more concretely, $\p$ is the Fourier transform of $\x$. Indeed, we can transform between $\x$ and $\p$ with the special rotation gate $F:=R(\frac{\pi}{2})$. Using a functional representation, the $\x$ operator has the effect of multiplication $\x\psi(x) = x \psi(x)$. In this same representation, $\p$ is proportional to the derivative operator, $\p \psi(x) = -i\frac{\partial}{\partial x}\psi(x)$, as expected from the theory of Fourier transforms.

Universality of the CV model is defined as the ability to approximate arbitrary transformations of the form
\begin{equation}
 U_H = \exp(-itH),
\end{equation}
where the generator $H=H(\mathbf{\x},\mathbf{\p})$ is a polynomial function of $(\mathbf{\x}, \mathbf{\p})$ with arbitrary but fixed degree \cite{lloyd1999quantum}. Crucially, such transformations are unitary in the Hilbert space picture, but can have complex nonlinear effects in the phase space picture, a fact that we later make use of for designing quantum neural networks. A set of gates is universal if it can be used to build any $U_H$ through a polynomial-depth quantum circuit. 
In fact, a universal gate set for CV quantum computing consists of the following ingredients: all the Gaussian transformations from Eq. (\ref{eq:rotation})-(\ref{eq:beamsplitter}), combined with any single non-Gaussian transformation, which corresponds to a nonlinear function on the phase space variables $(\bx,\bp)$. 
This is analogous to classical neural networks, where affine transformations combined with a single class of nonlinearity are sufficient to universally approximate functions. Commonly encountered non-Gaussian gates are the cubic phase gate $V(\gamma) = \exp(i\tfrac{\gamma}{3}\x^3)$ and the Kerr gate $K(\kappa) = \exp(i\kappa\hat{n}^2)$.

\section{Continuous-variable quantum neural networks}
\label{sec:qnn}

\begin{figure}[t]
\input{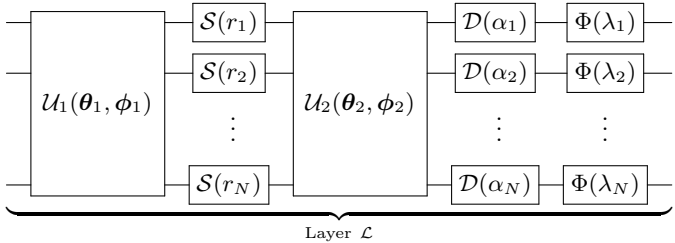}
\caption{The circuit structure for a single layer of a CV quantum neural network: an interferometer, local squeeze gates, a second interferometer, local displacements, and finally local non-Gaussian gates. The first four components carry out an affine transformation, followed by a final nonlinear transformation.}
\label{fig:layer_circuit}
\end{figure}

In this section, we present a scheme for quantum neural networks using the CV framework. It is inspired from two sides. First, from the structure of classical neural networks, which are universal function approximators and have demonstrated impressive performance on many practical problems. Second, from variational quantum circuits, which have recently become the predominant way of thinking about algorithms on near-term quantum devices \cite{peruzzo2014variational,moll2017quantum,verdon2017quantum,farhi2018classification,schuld2018quantum,schuld2018circuit,dallaire2018quantum, benedetti2018adversarial, havlicek2018supervised}. The main idea is the following: the fully connected neural network architecture provides a powerful and intuitive ansatz for designing variational circuits in the CV model. 

We will first introduce the most general form of the quantum neural network, which is the analogue of a classical fully connected network. We then show how a classical neural network can be embedded into the quantum formalism as a special case (where no superposition or entanglement is created), and discuss the universality and computational complexity of the fully quantum network.  
As modern deep learning has moved beyond the basic feedforward architecture, considering ever more specialized models, we will also discuss how to extend or specialize the quantum neural network to various other cases, specifically recurrent, convolutional, and residual networks. In Table \ref{tab:correspond}, we give a high-level matching between neural network concepts and their CV analogues. 

\begin{figure}[t]
\vspace{-0.1cm}
$$\centerline{
 \Qcircuit @C=1em @R=0.5em {
 & \multigate{7}{\mathcal{L}_1} & \multigate{7}{\mathcal{L}_2} & \multigate{7}{\mathcal{L}_3} & \\
 & \ghost{\mathcal{L}_1} & \ghost{\mathcal{L}_2} & \ghost{\mathcal{L}_3} & \\
 & \ghost{\mathcal{L}_1} & \ghost{\mathcal{L}_2} & \ghost{\mathcal{L}_3} & \\
 & \ghost{\mathcal{L}_1} & \ghost{\mathcal{L}_2} & \ghost{\mathcal{L}_3} & \\
 & \ghost{\mathcal{L}_1} & \ghost{\mathcal{L}_2} & \ghost{\mathcal{L}_3} & \multigate{3}{\mathcal{L}_4} & \\
 & \ghost{\mathcal{L}_1} & \ghost{\mathcal{L}_2} & \ghost{\mathcal{L}_3} & \ghost{\mathcal{L}_4} & \\
 & \ghost{\mathcal{L}_1} & \ghost{\mathcal{L}_2} & \ghost{\mathcal{L}_3} & \ghost{\mathcal{L}_4} & \multigate{1}{\mathcal{L}_5} & \\
 & \ghost{\mathcal{L}_1} & \ghost{\mathcal{L}_2} & \ghost{\mathcal{L}_3} & \ghost{\mathcal{L}_4} & \ghost{\mathcal{L}_5} & \gate{\mathcal{L}_6} & \meter \\
}
}$$
\caption{An example multilayer continuous-variable quantum neural network. In this example, the later layers are progressively decreased in size. Qumodes can be removed either by explicitly measuring them or by tracing them out. The network input can be classical, e.g., by displacing each qumode according to data, or quantum. The network output is retrieved via measurements on the final qumode(s).}
\label{fig:multilayer_circuit}
\end{figure}
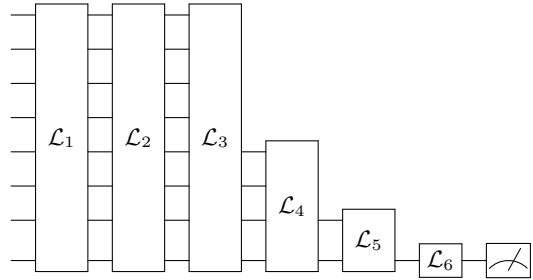

\subsection{Fully connected quantum layers}

A general CV quantum neural network is built up as a sequence of layers, with each layer containing every gate from the universal gate set. Specifically, a layer $\mathcal{L}$ consists of the successive gate sequence shown in Fig. \ref{fig:layer_circuit}: 
\begin{equation}
\label{eq:quantum_layer}
\mathcal{L}:= \Phi \circ \mathcal{D} \circ \mathcal{U}_2 \circ \mathcal{S} \circ \mathcal{U}_1,
\end{equation}
where $\mathcal{U}_i = \mathcal{U}_i(\bm{\theta},\bm{\phi})$ are general $N$-port linear optical interferometers containing beamsplitter and rotation gates, $\mathcal{D}=\otimes_{i=1}^N D(\alpha_i)$ and $\mathcal{S}=\otimes_{i=1}^N S(r_i)$ are collective displacement and squeezing operators (acting independently on each mode) and $\Phi = \Phi(\bm{\lambda})$ is some non-Gaussian gate, e.g., a cubic phase or Kerr gate. The collective gate variables $(\bm{\theta}, \bm{\phi}, \bm{r}, \bm{\alpha}, \bm{\lambda})$ form the free parameters of the network, where $\bm{\lambda}$ can be optionally kept fixed.

The sequence of Gaussian transformations $\mathcal{D} \circ \mathcal{U}_2 \circ \mathcal{S} \circ \mathcal{U}_1$ is sufficient to parameterize every possible unitary affine transformation on $N$ qumodes. In the phase space picture, this corresponds to the transformation of Eq. (\ref{eq:gen_gaussian_transf}). This sequence thus has the role of a `fully connected' matrix transformation. Interestingly, adding a nonlinearity uses the same component that adds universality: a non-Gaussian gate $\Phi$. 
Using $\mathbf{z}=(\bx, \bp)$, we can write the combined transformation in a form reminiscent of Eq. (\ref{eq:single_layer}), namely
\begin{equation}
  \mathcal{L}(\mathbf{z}) = \Phi(M\mathbf{z} + \bm{\alpha}).
\end{equation}
Thanks to the CV encoding, we get a nonlinear functional transformation while still keeping the quantum circuit unitary.

Similar to the classical setup, we can stack multiple layers of this type end-to-end to form a deeper network (Fig. \ref{fig:multilayer_circuit}). The quantum state output from one layer is used as the input for the next. 
Different layers can be made to have different widths by adding or removing qumodes between layers. Removal can be accomplished by measuring or tracing out the extra qumodes. In fact, conditioning on measurements of the removed qumodes is another method for performing non-Gaussian transformations \cite{andersen2015hybrid}.
This architecture can also accept classical inputs. We can do this by fixing some of the gate arguments to be set by classical data rather than free parameters, for example by applying a displacement $\mathcal{D}(\bx)$ to the vacuum state to prepare the state $\mathcal{D}(\bx)\ket{\bm{0}}$. This scheme can be thought of as an embedding of classical data into a quantum feature space \cite{schuld2018quantum}.
The output of the network can be obtained by performing measurements and/or computing expectation values. The choice of measurement operators is flexible; different choices (homodyne, heterodyne, photon-counting, etc.) may be better suited for different situations.

\subsection{Embedding classical neural networks}
\label{ssec:embedding}

The above scheme for a CV quantum neural network is quite flexible and general. 
In fact, it includes classical neural networks as a special case, where we don't create any superposition or entanglement. 
We now present a mathematical recipe for embedding a classical neural network into the quantum CV formalism. We give the recipe for a single feedforward layer; multilayer networks follow straightforwardly. 
Throughout this part, we will represent $N$-dimensional real-valued vectors $\bx$ using $N$-mode quantum optical states built from the eigenstates $\ket{x_i}$ of the operators $\x_i$:
\begin{equation}
 \bx\leftrightarrow\ket{\bx}:=\ket{x_1}\otimes\dots\otimes\ket{x_N}. 
\end{equation}
For the first layer in a network, we create the input $\bx$ by applying the displacement operator $\mathcal{D}(\bx)$ to the state $\ket{\bx=\mathbf{0}}$. Subsequent layers will use the output of the previous layer as input. To read out the output from the final layer, we can use ideal homodyne detection in each qumode, which projects onto the states $\ket{x_i}$ \cite{serafini2017quantum}.

We would like to enact a fully connected layer (Eq. (\ref{eq:single_layer})) completely within this encoding, i.e., 
\begin{equation}
  \ket{\bx}\mapsto \ket{\varphi(W\bx + \mathbf{b})}. 
\end{equation}
This transformation will take place entirely within the $\bx$ coordinates; we will not use the momentum variables. We thus want to restrict our quantum network to never mix between $\mathbf{\x}$ and $\mathbf{\p}$.
To proceed, we will break the overall computation into separate pieces. Specifically, we split up the weight matrix using a singular value decomposition, $W = {O}_2 {\Sigma} {O}_1$, where the ${O}_k$ are orthogonal matrices and ${\Sigma}$ is a positive diagonal matrix. For simplicity, we assume that $W$ is full rank. Rank-deficient matrices form a measure-zero subset in the space of weight matrices, which we can approximate arbitrarily closely with full-rank matrices.

\paragraph*{Multiplication by an orthogonal matrix.} The first step in Eq. (\ref{eq:quantum_layer}) is to apply an interferometer $\mathcal{U}_1$, which corresponds to the rightmost orthogonal matrix $K_1$ in Eq. (\ref{eq:bloch_messiah}). In order not to mix $\mathbf{\x}$ and $\mathbf{\p}$, we must restrict to block-diagonal $K_1$. With respect to Eqs. (\ref{eq:symplectic_orthogonal_block})-(\ref{eq:symplectic_orthogonal_eqns2}), this means that $C$ is an orthogonal matrix and $D=0$. 
This choice corresponds to an interferometer which only contains phaseless beamsplitters.
With this restriction, we have 
\begin{align}
 \label{eq:passive_interferometer_effect}
 \mathcal{U}_1\ket{\bx} & = \mathcal{U}_1\Bigg[\bigotimes_{i=1}^N\ket{x_i}\Bigg] \nonumber \\
 & = \bigotimes_{i=1}^N \Bigg| \sum_{j=1}^N C_{ij} x_j \Bigg\rangle \nonumber \\
 & = \ket{C\bx}.
\end{align}
The full derivation of this expression can be found in Appendix \ref{sec:interferometer_proof}.
Thus, the phaseless linear interferometer $\mathcal{U}_1$ is equivalent to multiplying the encoded data by an orthogonal matrix $C$. To connect to the weight matrix $W=O_1\Sigma O_2$, we choose the interferometer which has $C=O_1$. 
A similar result holds for the other interferometer $\mathcal{U}_2$. 

\paragraph*{Multiplication by a diagonal matrix.} For our next element, consider the squeezing gate. 
The effect of squeezing on the $\x_i$ eigenstates is \cite{kok2010introduction}
\begin{equation}
  S(r_i)\ket{x_i} = \sqrt{c_i}\ket{c_i x_i},
\end{equation}
where $c_i=e^{-r_i}$. An arbitrary positive scaling $c_i$ can thus be achieved by taking $r_i=\log(c_i)$. Note that squeezing leads to compression (positive $r_i$, $c_i\leq 1$), while antisqueezing gives expansion (negative $r_i$, $c_i \geq 1$), matching with Eq. (\ref{eq:squeezing}). 
A collection of local squeezing transformations thus corresponds to an elementwise scaling of the encoded vector, 
\begin{equation}
  \mathcal{S}(\mathbf{r})\ket{\bx} = e^{-\tfrac{1}{2}\sum_i r_i}\ket{\Sigma\bx},
\end{equation}
where $\Sigma:=\mathrm{diag}(\{c_i\})> 0$. We note that since the $\ket{x_i}$ eigenstates are not normalizable, the prefactor has limited formal consequence.

\paragraph*{Addition of bias.} Finally, it is well-known that the displacement operator acting locally on quadrature eigenstates has the effect
\begin{equation}
 \mathcal{D}(\alpha_i)\ket{x_i} = \ket{x_i + \alpha_i},
\end{equation}
for $\alpha_i\in\mathbb{R}$, which collectively gives
\begin{equation}
  \mathcal{D}(\bm{\alpha})\ket{\bx} = \ket{\bx + \bm{\alpha}}.
\end{equation}
Thus, to achieve a bias translation of $\mathbf{d}$, we can simply displace by $\bm{\alpha}=\mathbf{d}$.

\paragraph*{Affine transformation.} Putting these ingredients together, we have 
\begin{align}
 \mathcal{D} \circ \mathcal{U}_2 \circ \mathcal{S} \circ \mathcal{U}_1 \ket{\bx} & \propto \ket{O_2\Sigma O_1\bx + \bm{d}} \nonumber \\
 & = \ket{W\bx+\bm{d}},\label{affine}
\end{align}
where we have omitted the parameters for clarity.
Hence, using only Gaussian operations which do not mix $\bx$ and $\bp$, we can effectively perform arbitrary full-rank affine transformations amongst the vectors $\ket{\bx}$.

\paragraph*{Nonlinear function.} 
To complete the picture, 
we need to find a non-Gaussian transformation $\Phi$ which has the following effect
\begin{equation}
  \Phi\ket{\bx} = \ket{\varphi(\bx)},
\end{equation}
where $\varphi:\mathbb{R}\rightarrow\mathbb{R}$ is some nonlinear function. We will restrict to an element-wise function, i.e., $\Phi$ acts locally on each mode, similar to the activation function of a classical neural network. 
For simplicity, we will consider $\varphi$ to be a polynomial of fixed degree. By allowing the degree of $\varphi$ to be arbitrarily high, we can approximate any function which has convergent Taylor series. The most general form of a quantum channel consists of appending an ancilla system, performing a unitary transformation on the combined system, and tracing out the ancilla. For qumode $i$, we will append an ancilla $i'$ in the $x=0$ eigenstate, i.e.,
\begin{equation}
  \ket{x}_i \mapsto \ket{x}_i \ket{0}_{i'},
\end{equation}
where, for clarity, we have made the temporary notational change $\ket{x_i}\leftrightarrow\ket{x}_i$.

Consider now the unitary $V_\varphi:=\exp{(i \varphi(\x_i) \otimes \p_{i'})}$, where $\varphi(\x_i)$ is understood as a Taylor series using powers of $\x_i$. 
Applying this to the above two-mode system, we get
\begin{align}
 \exp{(-i \varphi(\x_i) \otimes \p_{i'})}\ket{x}_i \ket{0}_{i'} = & \exp{(-i \varphi(x_i) \p_{i'})}\ket{x}_i \ket{0}_{i'} \nonumber \\
                                                 = & \mathcal{D}_{i'}{(\varphi(x_i)})\ket{x}_i \ket{0}_{i'} \nonumber \\
                                                 = & \ket{x}_i \ket{\varphi(x)}_{i'},
\label{eq:nonlinearity}                                                 
\end{align}
where we have recognized that $\p$ is the generator of displacements in $x$. We can now swap modes $i$ and $i'$ (using a perfectly reflective beamsplitter) and trace out the ancilla. The combined action of these operations leads to the overall transformation
\begin{equation}
  \ket{x_i} \mapsto \ket{\varphi(x_i)}.
\end{equation}
Alternatively, we are free to keep the system in the form $\ket{x_i}\ket{\varphi(x_i)}$; this can be useful for creating residual quantum neural networks.

Together, the above sequence of Gaussian operations, followed by a non-Gaussian operation, lead to the desired transformation $\ket{\bx}\mapsto\ket{\varphi(W\bx +\mathbf{b})}$, which is the same as a single-layer classical neural network. We remark finally that the states $\ket{x}$ were used in order to provide a convenient mathematical embedding; in a practical CV device, we would need to approximate the states $\ket{x}$ via finitely squeezed states. 
In practice, the general quantum neural network framework does not require any particular choice of basis or encoding. 
Because of this additional flexibility, the full quantum network has larger representational capacity than a conventional neural network and cannot be efficiently simulated by classical models, as we now discuss.

\subsection{The power of CV neural networks}

None of the transformations considered in the previous section ever generate superpositions or entanglement. 
A distinguishing feature of quantum physics is that we can act not only on some fixed basis states, e.g., the states $\ket{\bx}$, but also on superpositions -- that is, linear combinations -- of those basis states, $\ket{\psi} = \int \psi(\bx)\ket{\bx}d\bx$, where $\psi(\bx)$ is a multimode wavefunction. 
The general CV neural network provides greater freedom in the allowed operations by leveraging the power of universal quantum computation. Indeed, the quantum gates in a single layer form a universal gate set, which implies that a CV quantum neural network shares all the capabilities of a universal CV quantum computer. 

To see this, consider an arbitrary quantum computation and its decomposition in terms of a circuit consisting of a sequence of gates from universal gate set. We assign a quantum neural network to this circuit by replacing each gate in the circuit by a single layer. Since each layer contains all gates from the universal set, it can reproduce the action of the single selected gate by setting the parameters of all other gates to zero. Therefore the full network can also replicate the complete quantum circuit.

Since CV quantum neural networks are capable of universal CV quantum computation, in general we do not expect that they can be efficiently simulated on a classical computer. This statement can be put on firmer ground by considering a simple modification to the classical neural network embedding from Sec. \ref{ssec:embedding}. Specifically, we carry out a Fourier transform on all modes at the beginning and end of the network. The result is that input states $\ket{\bx}$ are replaced by momentum eigenstates $\ket{\bp}$ and the position homodyne measurements are replaced with momentum homodyne measurements. A momentum eigenstate is an equal superposition over all position eigenstates and thus this circuit can be interpreted as acting on an equal superposition of all classical inputs.

The resulting circuits, consisting of input momentum eigenstates, a unitary transformation that is diagonal in the position basis, and momentum homodyne measurements, are known as continuous-variable instantaneous quantum polynomial (CV-IQP) circuits. It was proven in Ref. \cite{douce2017iqp} that efficient exact classical simulation of CV-IQP circuits would imply a collapse of the polynomial hierarchy to third level. This result was extended in Ref. \cite{arrazola2017quantum} to the case of approximate classical simulation, under the validity of a plausible conjecture concerning the computational complexity of evaluating high-dimensional integrals. Thus, even a simple modification of the classical embedding presented above gives quantum neural networks the ability to perform tasks that would require exponentially many resources to replicate on classical devices.

\subsection{Beyond the fully connected architecture}

\begin{figure*}[t]
\vspace{0.25cm}
\includegraphics[width=1.0 \textwidth]{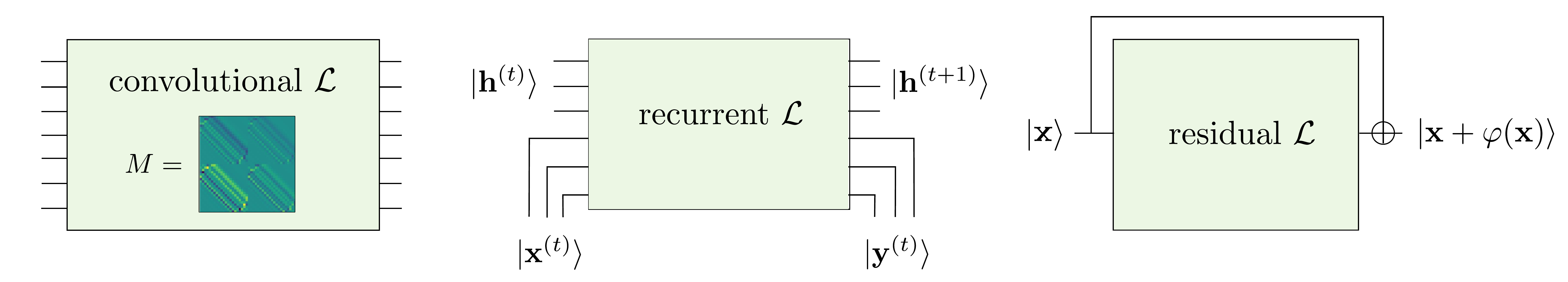}
\caption{Quantum adaptations of the convolutional layer, recurrent layer, and residual layer. The convolutional layer is enacted using a Gaussian unitary with translationally invariant Hamiltonian, resulting in a corresponding symplectic matrix that has a block Toeplitz structure. The recurrent layer combines an internal signal from previous layers with an external source, while the residual layer combines its input and output signals using a controlled-X gate.}
\label{Fig:NNs}
\end{figure*}

Modern deep learning techniques have expanded beyond the basic fully connected architecture. Powerful deep learning software packages \cite{bergstra2010theano, jia2014caffe, maclaurin2015autograd, abadi2016tensorflow, paszke2017automatic} have allowed researchers to explore more specialized networks or complicated architectures. For the quantum case, we should also not feel restricted to the basic network structure presented above. Indeed, the CV model gives us flexibility to encode problems in a variety of representations. For example, we can use the phase space picture, the wavefunction picture, the Hilbert space picture, or some hybrid of these. We can also encode information in coherent states, squeezed states, Fock states, or superpositions of these states. 
Furthermore, by choosing the gates and parameters to have particular structure, we can specialize our network ansatz to more closely match a particular class of problems. This can often lead to more efficient use of parameters and better overall models. In the rest of this section, we will highlight potential quantum versions of various special neural network architectures; see Fig.~\ref{Fig:NNs} for a visualization.

\paragraph*{Convolutional network.}
A common architecture in classical neural networks is the convolutional network, or \emph{convnet} \cite{lecun1989backpropagation}. Convnets are particularly well-suited for computer vision and image recognition problems because they reflect a simple yet powerful observation: since the task of detecting an object is largely independent of where the object appears in an image, the network should be equivariant to translations \cite{goodfellow2016deep}. Consequently, the linear transformation $W$ in a convnet is not fully connected; rather, it is a specialized sparse linear transformation, namely a convolution. In particular, for one-dimensional convolutions, the matrix $W$ has a \emph{Toeplitz} structure, with entries repeated along each diagonal.
This is similar to the well-known principle in physics that symmetries in a physical system can lead to simplifications of our physical model for that system (e.g., Bloch's Theorem \cite{bloch1929quantenmechanik} or Noether's Theorem \cite{noether1918invariante}). 

We can directly enforce translation symmetry on a quantum neural network model by making each layer in the quantum circuit translationally invariant. Concretely, consider the generator $H=H(\mathbf{\x}, \mathbf{\p})$ of a Gaussian unitary, $\mathcal{U}=\exp(-itH)$. Suppose that this generator is translationally invariant, i.e., $H$ does not change if we map $(\x_i,\p_i)$ to $(\x_{i+1},\p_{i+1})$. Then the symplectic matrix $M$ that results from this Gaussian unitary will have the form 
\begin{equation}
 M =  
   \begin{bmatrix}
     M_{\bx\bx} & M_{\bx\bp} \\
     M_{\bp\bx} & M_{\bp\bp}
   \end{bmatrix},
\end{equation}
where each $M_{\mathbf{u}\mathbf{v}}$ is itself a Toeplitz matrix, i.e., a one-dimensional convolution (see Appendix \ref{app:convolutional_proof}). The matrix $M$ can be seen as a special kind of convolution that respects the uncertainty principle: performing a convolution on the $\bx$ coordinates naturally leads to a conjugate convolution involving $\bp$.
The connection between translationally invariant Hamiltonians and convolutional networks was also noted in \cite{lin2017does}.

\paragraph*{Recurrent network.}
This is a special-purpose neural network which is used widely for problems involving sequences \cite{graves2012supervised}, e.g., time series or natural language. 
A recurrent network can be pictured as a model which takes two inputs for every time step $t$. One of these inputs, $\bx^{(t)}$, is external, coming from a data source or another model. The other input is an internal state $\mathbf{h}^{(t)}$, which comes from the same network, but at a previous time-step (hence the name recurrent). These inputs are processed through a neural network $f_{\bm{\theta}}(\bx^{(t)}, \mathbf{h}^{(t)})$, and an output $\mathbf{y}^{(t)}$ is (optionally) returned.
Similar to a convolutional network, the recurrent architecture encodes translation symmetry into the weights of the model. However, instead of spatial translation symmetry, recurrent models have time translation symmetry. In terms of the network architecture, this means that the model reuses the same weights matrix $W$ and bias vector $b$ in every layer. In general, $W$ or $b$ are unrestricted, though more specialized architectures could also further restrict these.

This architecture generalizes straightforwardly to quantum neural networks, with the inputs, outputs, and internal states employing any of the data-encoding schemes discussed earlier. It is particularly well-suited to an optical implementation, since we can connect the output modes of a quantum circuit back to the input using optical fibres. This allows the same quantum optical circuit to be reused several times for the same model. We can reserve a subset of the modes for the data input and output channels, with the remainder used to carry forward the internal state of the network between time steps.

\begin{figure*}
\includegraphics[width=\textwidth]{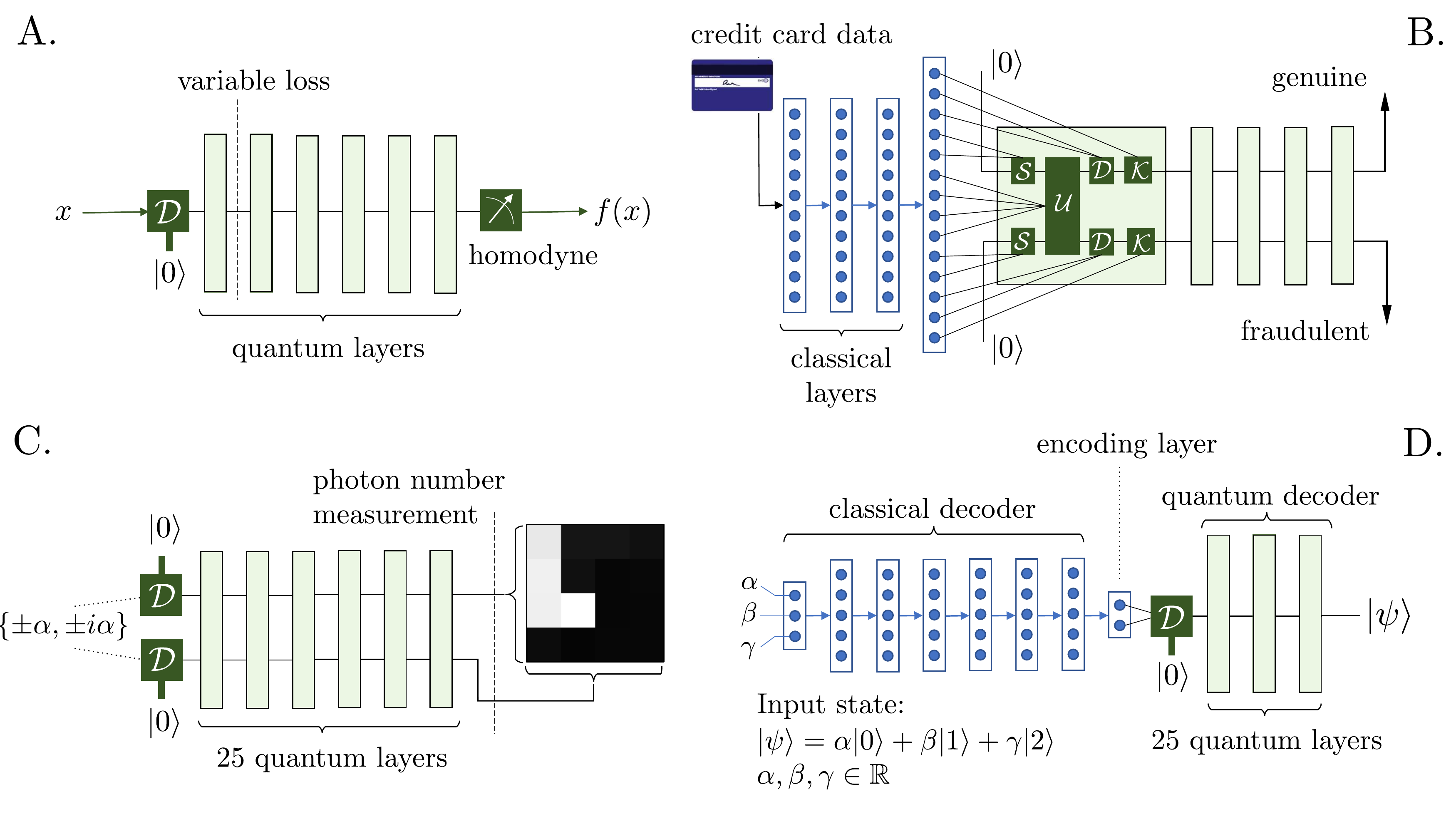}
\caption{Machine learning problems and architectures explored in this work: A. curve fitting of functions $f(x)$ is achieved through a multilayer network, with $x$ encoded through a position displacement on the vacuum and $f(x)$ through a position homodyne measurement at output; B. credit card fraud detection using a hybrid classical-quantum classifier, with the classical network controlling the parameters of an input layer; C. image generation of the Tetris dataset from input displacements to the vacuum, with output image encoded in photon number measurements at the output mode; D. hybrid classical-quantum autoencoder for finding a continuous phase-space encoding for the first three Fock states.}\label{Fig:Architectures}
\end{figure*}

\paragraph*{Residual network.}
The residual network \cite{he2016deep}, or \emph{resnet}, is a more recent innovation than the convolutional and recurrent networks. While these other models are special cases of feedforward networks, the resnet uses a modified network topology. Specifically, `shortcut connections,' which perform a simple identity transformation, are introduced between layers. Using these shortcuts, the output of a layer can be added to its input. If a layer by itself would perform the transformation $\mathcal{F}$, then the corresponding residual network performs the transformation
\begin{equation}
 \bx \mapsto \bx + \mathcal{F}(\bx).
\end{equation}

To perform residual-type computation in a quantum neural network, we look back to Eq. (\ref{eq:nonlinearity}), where a two-mode unitary was given which carries out the transformation 
\begin{equation}
 \label{eq:residual_part1}
 \ket{x}\ket{0} \mapsto \ket{x}\ket{\varphi(x)},
\end{equation}
where $\varphi$ is some desired non-Gaussian function. To complete the residual computation, we need to sum these two values together. This can be accomplished using the controlled-X (or $SUM$) gate $C_X$ \cite{gottesman2001encoding}, which can be carried out with purely Gaussian operations, namely squeezing and beamsplitters \cite{strawberryfields_cxgate}. 
Adding a $C_X$ gate after the transformation in Eq. (\ref{eq:residual_part1}), we obtain
\begin{equation}
 \ket{x}\ket{0} \mapsto \ket{x}\ket{x+\varphi(x)},
\end{equation}
which is a residual transformation. 
This residual transformation can also be carried out on arbitrary wavefunctions $\psi(x)$ in superposition, giving the general mapping
\begin{equation}
 \int \psi(x) \ket{x} dx \mapsto \int \psi(x) \ket{x}\ket{x + \varphi(x)} dx.
\end{equation}

\section{Numerical Experiments}
\label{sec:numerics}

We showcase the power and versatility of CV quantum neural networks by employing them in a range of machine learning tasks. The networks are numerically simulated using the Strawberry Fields software platform \cite{killoran2018strawberry} and the Quantum Machine Learning Toolbox app which is built on top of it. We use both automatic differentiation with respect to the quantum gate parameters, which is built into Strawberry Fields' TensorFlow \cite{abadi2016tensorflow} quantum circuit simulator, as well as numerical algorithms to train these networks. Automatic differentiation techniques allow for a direct use of established optimization algorithms based on stochastic gradient descent. On the other hand, numerical techniques such as the finite-difference method or Nelder-Mead will allow training of hardware-based implementations of quantum neural networks.

\begin{center}
\begin{figure*}[t!]
\begin{tabular}{ccc}
\includegraphics[width=0.66 \columnwidth]{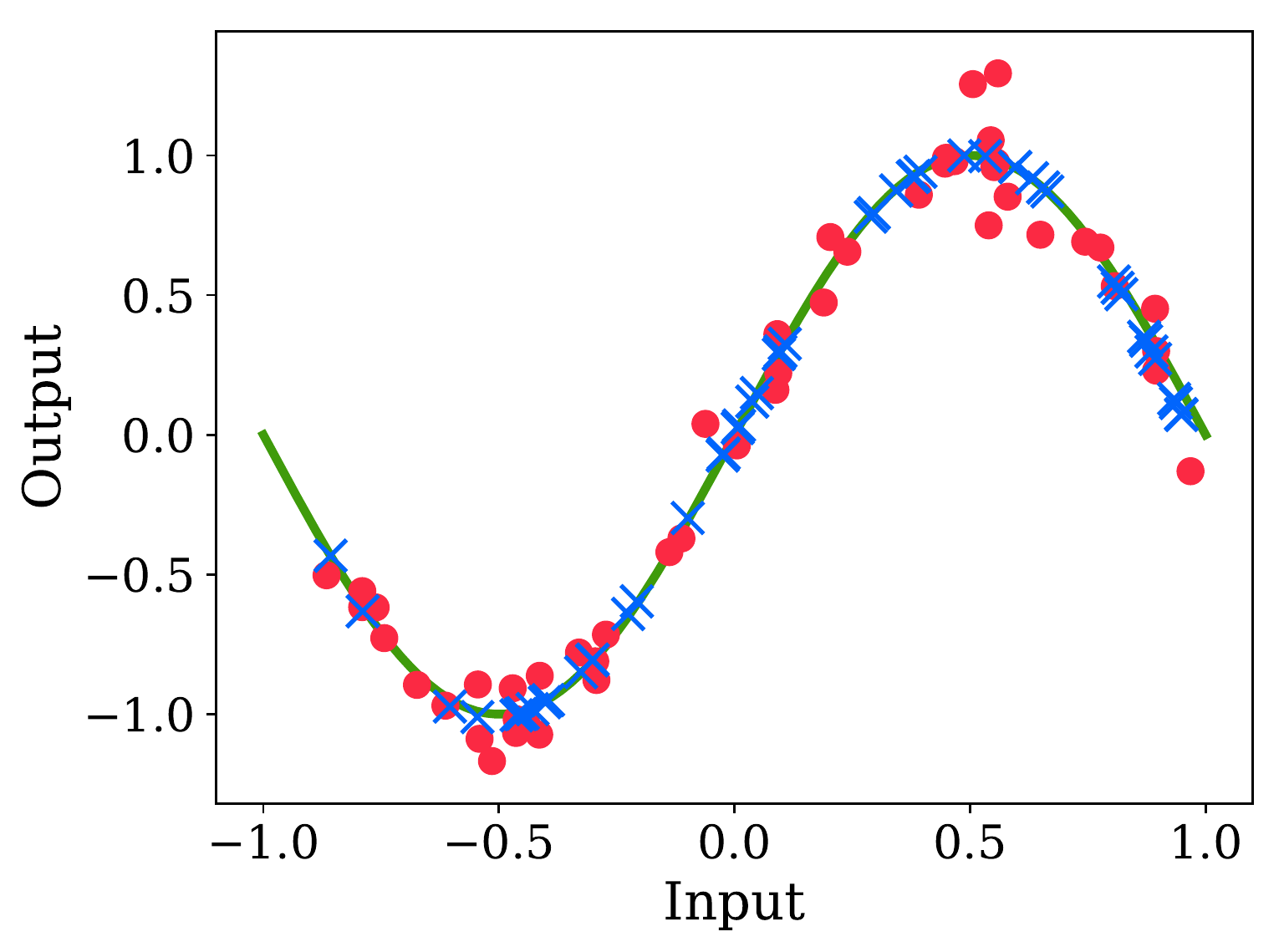}& 
\includegraphics[width=0.66 \columnwidth]{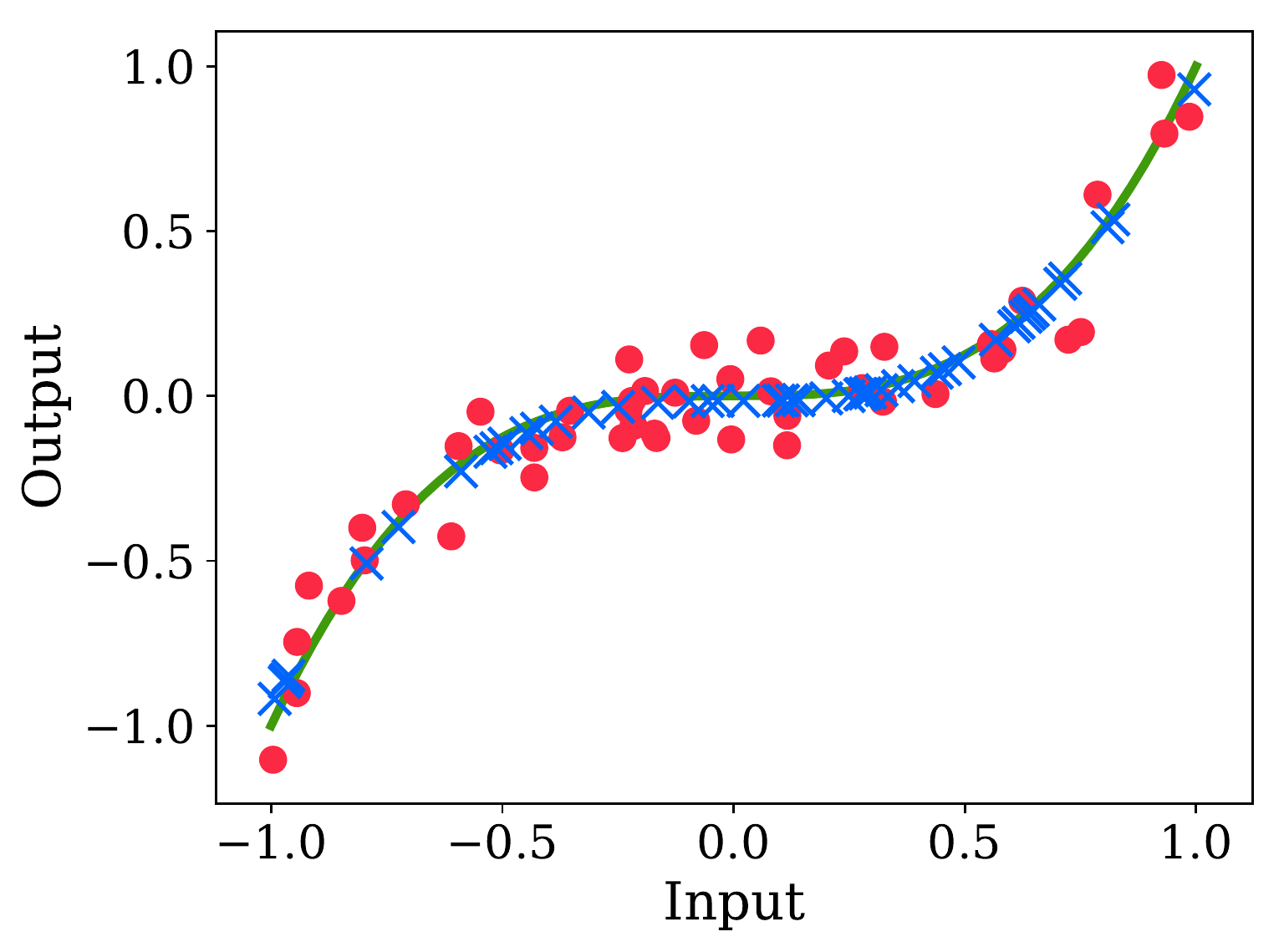}&
\includegraphics[width=0.66 \columnwidth]{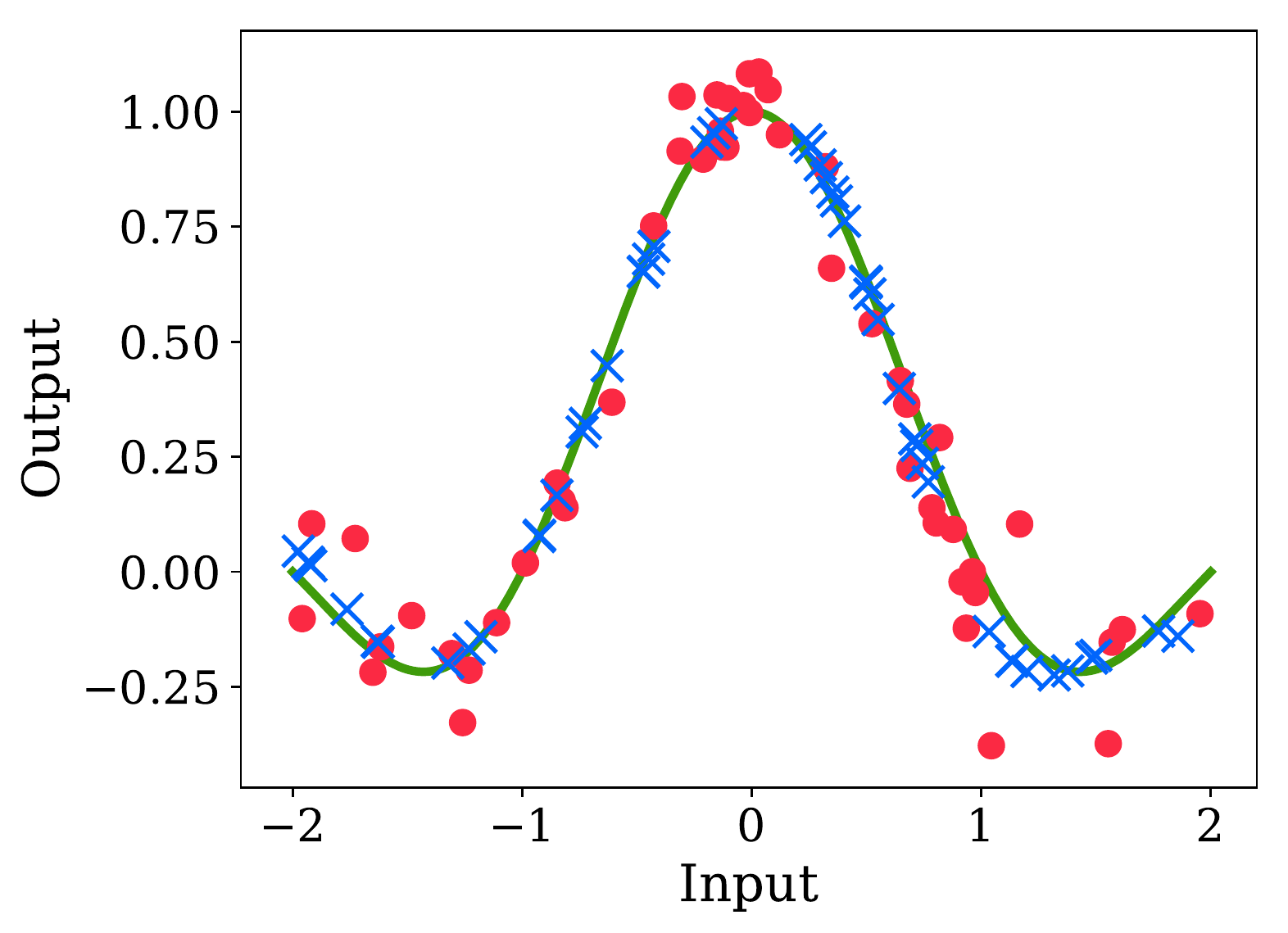}
\end{tabular}
\caption{Experiment A. Curve fitting with continuous-variable quantum neural networks. The networks consist of six layers and were trained for 2000 steps with a Hilbert-space cutoff dimension of 10. As examples, we consider noisy versions of the functions $\sin(\pi x)$, $x^3$, and $\text{sinc}(\pi x)$, displayed respectively from left to right. We set a standard deviation of $\epsilon=0.1$ for the noise. The training data is shown as red circles. The outputs of the quantum neural network for the test inputs are shown as blue crosses. The outputs of the circuit very closely resemble the noiseless ground truth curves, shown in green.}
\label{Fig:curvefitting}
\end{figure*}
\end{center}

We study several tasks in both supervised and unsupervised settings, with varying degrees of hybridization between quantum and classical neural networks. Some cases employ both classical and quantum networks whereas others are fully quantum. The architectures used are illustrated in Fig.~\ref{Fig:Architectures}. Unless otherwise stated, we employ the Adam optimizer~\cite{kingma2014adam} to train the networks and we choose the Kerr gate $K(\kappa)=\exp(i\kappa\hat{n}^2)$ as the non-Gaussian gate in the quantum networks. Our results highlight the wide range of potential applications of CV quantum neural networks, which will be further enhanced when deployed on dedicated hardware which exceeds the current limitations imposed by classical simulations.

\subsection{Training quantum neural networks}\label{Sec:CurveFit}
A prototypical problem in machine learning is curve fitting: learning a given relationship between inputs and outputs. 
We will use this simple setting to analyze the behaviour of CV quantum neural networks with respect to different choices for the model architecture, cost function, and optimization algorithm. 
We consider the simple case of training a quantum neural network to reproduce the action of a function $f(x)$ on one-dimensional inputs $x$, when given a training set of noisy data. This is summarized in Fig.~\ref{Fig:Architectures}(a). We encode the classical inputs as position-displaced vacuum states $\mathcal{D}(x)\ket{0}$, where $\mathcal{D}(x)$ is the displacement operator and $\ket{0}$ is the single-mode vacuum. Let $\ket{\psi_x}$ be the output state of the circuit given input $\mathcal{D}(x)\ket{0}$. The goal is to train the network to produce output states whose expectation value for the quadrature operator $\hat{x}$ is equal to $f(x)$, i.e., to satisfy the relation $\bra{\psi_x}\hat{x}\ket{\psi_x} =f(x)$ for all $x$. 

To train the circuits, we use a supervised learning setting where the training and test data are tuples $(x_i,f(x_i))$ for values of $x_i$ chosen uniformly at random in some interval. We define the loss function as the mean square error (MSE) between the circuit outputs and the desired function values
\beq\label{Eq: loss function}
L = \frac{1}{N}\sum_{i=1}^N [f(x_i)-\bra{\psi_{x_i}}\hat{y}\ket{\psi_{x_i}}]^2.
\eeq
To test this approach in the presence of noise in the data, we consider functions of the form $\tilde{f}(x) = f(x)+\Delta f$ where $\Delta f$ is drawn from a normal distribution with zero mean and standard deviation $\epsilon$. The results of curve fitting on three noisy functions are illustrated in Fig. \ref{Fig:curvefitting}. 

\paragraph*{Avoiding overfitting.}

Ideally, the circuits will produce outputs that are smooth and do not overfit the noise in the data. CV quantum neural networks are inherently adept at achieving smoothness because quantum states that are close to each other cannot differ significantly in their expectation value with respect to observables. Quantitatively, H\"older's inequality states that for any two states $\rho$ and $\sigma$ it holds that
\beq
\left|\text{Tr}[(\rho-\sigma)X]\right|\leq \|\rho-\sigma\|_1\|X\|_\infty
\eeq 
for any operator $X$. This smoothness property of quantum neural networks is clearly seen in Fig. \ref{Fig:curvefitting}, where the input/output relationship of quantum circuits gives rise to smooth functions that are largely immune to the presence of noise, while still being able to generalize from training to test data. We found that no regularization mechanism was needed to prevent overfitting of the problems explored here.

\begin{center}
\begin{figure}[t!]
\includegraphics[width=0.75\columnwidth]{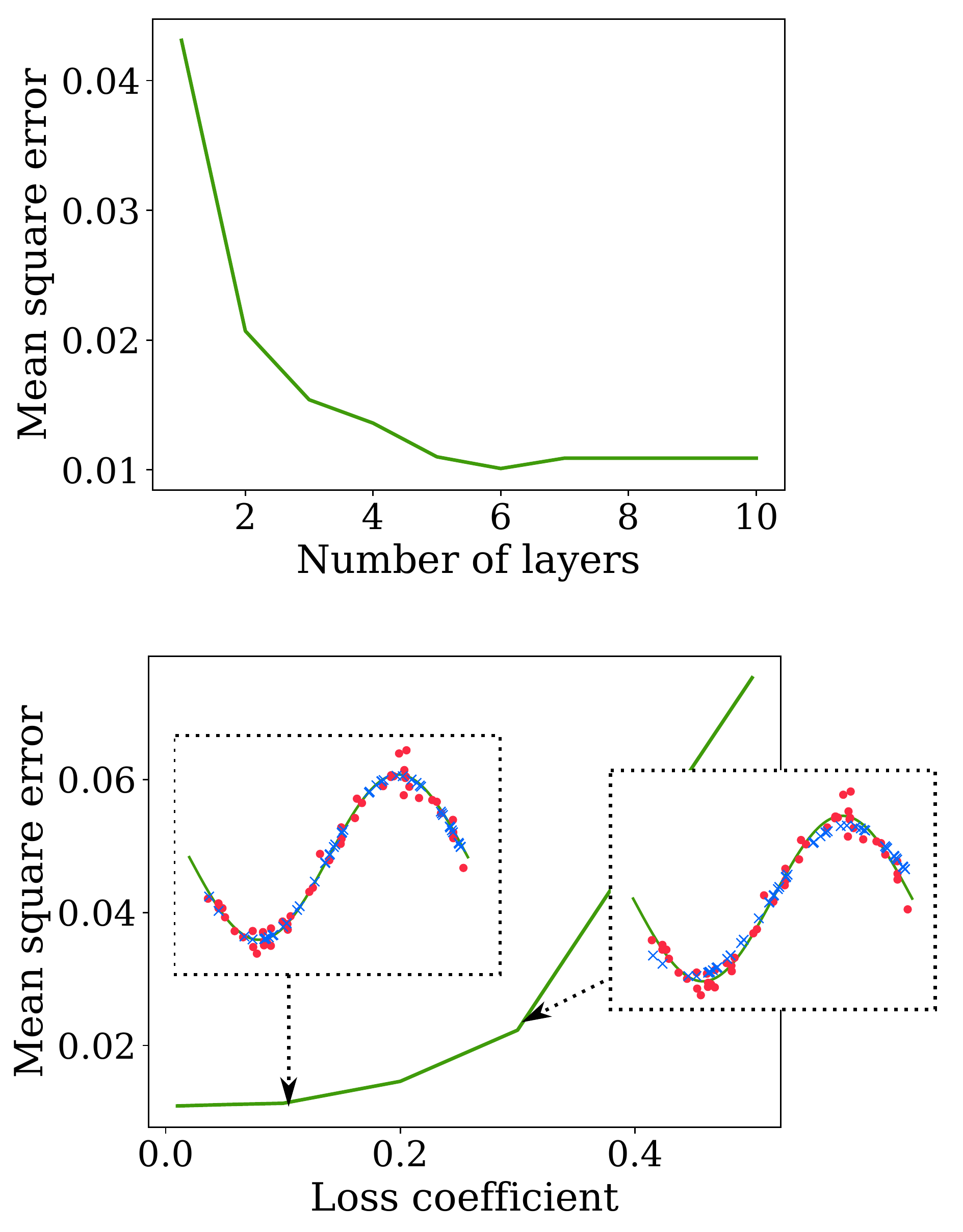}
\caption{MSE as a function of the number of layers and as a function of photon loss. The plots correspond to the task of fitting the function $\sin(\pi x)$ in the interval $x\in[-1,1]$. (Top) Increasing the number of layers is helpful until a saturation point is reached with six layers, after which little improvement is observed. (Bottom) The networks can be resilient to imperfections, as seen by the fact that only a slight deviation in the mean square error appears for losses of 10\% in each layer. The fits with a photon loss coefficient of 10\% and 30\% are shown in the inset.}\label{Fig:LayersLoss}
\end{figure}
\end{center}

\paragraph*{Improvement with depth.}
The circuit architecture is defined by the number of layers, i.e., the circuit depth. Fig. \ref{Fig:LayersLoss} (top) studies the effect of the number of layers on the final value of the MSE. A clear improvement for the curve fitting task is seen for up to six layers, at which point the improvements saturate. The MSE approaches the square of the standard deviation of the noise, $\epsilon^2 = 0.01$, as expected when the circuit is in fact reproducing the input-output relationship of the noiseless curve.

\paragraph*{Quantum device imperfections.}
We also study the effect of imperfections in the circuit, which for photonic quantum computers is dominated by photon loss. We model this using a lossy bosonic channel, with a loss parameter $\eta$. Here $\eta=0\%$ stands for perfect transmission (no photon loss). The lossy channel acts at the end of each individual layer, ensuring that the effect of photon loss increases with circuit depth. For example, a circuit with six layers and loss coefficient $\eta=10\%$ experiences a total loss of $46.9\%$.  The effect of loss is illustrated in Fig. \ref{Fig:LayersLoss} (bottom) where we plot the MSE as a function of $\eta$. The quality of the fit exhibits resilience to this imperfection, indicating that the circuit learns to compensate for the effect of losses.

\begin{figure}
 \begin{center}
\includegraphics[width=0.9\columnwidth]{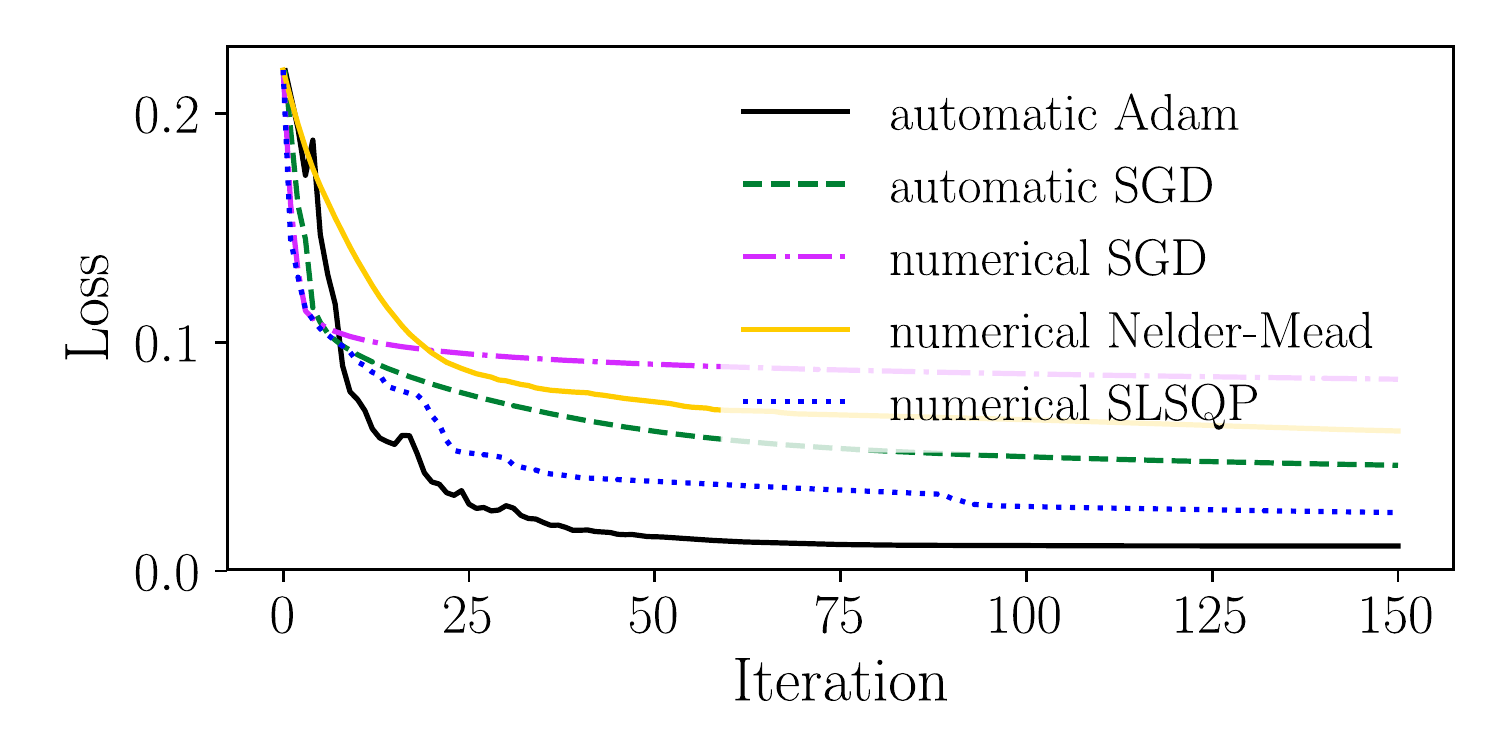}
\caption{Loss function for the different optimizers mentioned in the text.  
}
\label{Fig:Optimizers}
\end{center}  
\end{figure}

\paragraph*{Optimization methods.}
We also analyze different optimization algorithms for the sine curve-fitting problem. Fig. \ref{Fig:Optimizers} compares three numerical methods and two methods based on automatic differentiation. 
Numerical SGD approximates the gradients with a finite differences estimate. Nelder-Mead is a gradient-free technique, while the sequential least-squares programming (SLSQP) method solves quadratic subproblems with approximate gradients. These latter two converge significantly slower, but can have advantages in smoothness and speed per iteration. The Adam optimizer with adaptive learning rate performed better than vanilla SGD in this experiment.

\paragraph*{Penalties and regularization.}

In the numerical simulations of quantum circuits, each qumode is truncated to a given cutoff dimension in the infinite-dimensional Hilbert space of Fock states. During training, it is possible for the gate parameters to reach values such that the output states have significant support outside of the truncated Hilbert space. In the simulation, this results in unnormalized output states and unreliable computations. To address this issue, we add a penalty to the loss function that penalizes unnormalized quantum states. Given a set of output states $\{\ket{\psi_{x_i}}\}$, we define the penalty function
\beq
P(\{\ket{\psi_{x_i}}\})= \sum_i (|\bra{\psi_{x_i}}\Pi_{\mathcal{H}}\ket{\psi_{x_i}}|^2-1)^2,\label{Eq: Regularization}
\eeq
where $\Pi_{\mathcal{H}}$ is a projector onto the truncated Hilbert space of the simulation. This function penalizes unnormalized states whose trace is different to one. The overall cost function to be minimized is then
\beq
C=L+\gamma P(\{\ket{\psi_{x_i}}\}),
\eeq
where $\gamma>0$ is a user-defined hyperparameter. 

\begin{figure*}[t!]
 \begin{center}
\includegraphics[width=1.9\columnwidth]{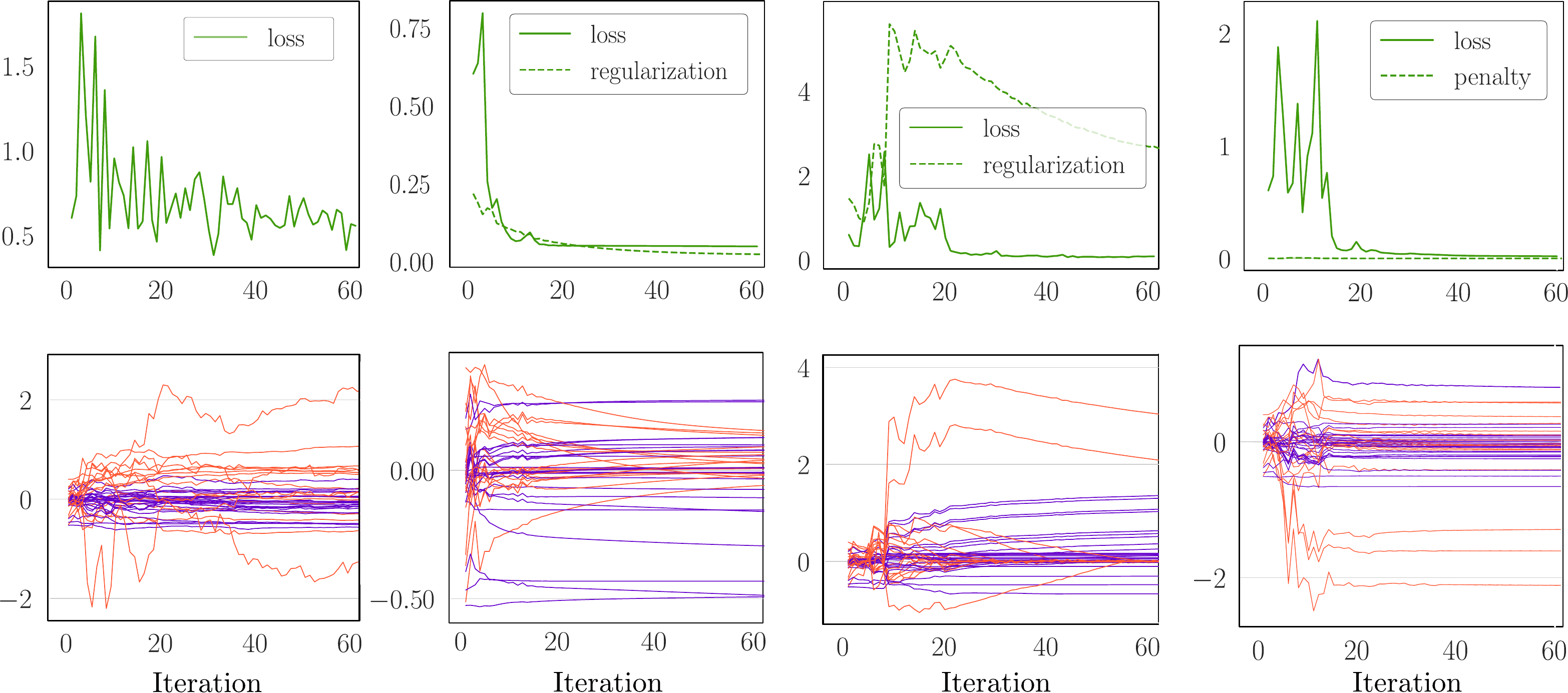}
\caption{Cost function and circuit parameters during 60 steps of stochastic gradient descent training for the task of fitting the sine function from Fig. \ref{Fig:LayersLoss}. The active parameters are plotted in orange, while all others are plotted in purple. As hyperparameters, we used an initial learning rate of $0.1$ which has an inverse decay of $0.25$, a penalty strength $\gamma=10$, a regularization strength of $0.5$, batch size of $50$, a cutoff of 10 for the Hilbert-space dimension, and randomly chosen but fixed initial circuit parameters.}
\label{Fig:Penalties}
\end{center}  
\end{figure*}

An alternate approach to the trace penalty is to regularize the circuit parameters that can alter the energy of the state, which we refer to as the active parameters. Fig. \ref{Fig:Penalties} compares optimizing the function of Eq. \eqref{Eq: Regularization} without any penalty (first column from the left), imposing an L2 regularizer (second column), using an L1 regularizer (third column), and using the trace penalty (fourth column). Without any strategy to keep the parameters small, learning fails due to unstable simulations: the trace of the state drops in fact to $0.1$. Both regularization strategies as well as the trace penalty manage to bring the loss function to almost zero within a few steps while maintaining the unit trace of the state. However, there are interesting differences. While L2 regularization decreases the magnitude of the active parameters, L1 regularization dampens all but two of them. The undamped parameters turn out to be the circuit parameters for the nonlinear gates in layer $3$ and $4$, a hint that these nonlinearities are most essential for the task. The trace penalty induces heavy fluctuations in the loss function for the first $20$ steps, but finds parameters that are larger in absolute value than those found by L2 regularization, with a lower final loss. 

\subsection{Supervised learning with hybrid networks}
\label{sec:fraud}

Classification of data is a canonical problem in machine learning. We construct a hybrid classical-quantum neural network as a classifier to detect fraudulent transactions in credit card purchases. In this hybrid approach, a classical neural network is used to control the gate parameters of the quantum network, the output of which determines whether the transactions are classified as genuine or fraudulent. This is illustrated in Fig.~\ref{Fig:Architectures}(b).

\paragraph*{Data preparation.}
For the experiment, data was taken from a publicly available database of labelled historical credit card transactions which are flagged as either \emph{fraudulent} or \emph{genuine}~\cite{dal2015calibrating}. The data is composed of $28$ features derived through a principal component analysis of the raw data, providing an anonymization of the transactions. Of the $284,807$ provided transactions, only $0.172 \%$ are fraudulent.
We create training and test datasets by splitting the fraudulent transactions in two and combining each subset with genuine transactions. For the training dataset, we undersample the genuine transactions by randomly selecting them so that they outnumber the fraudulent transactions by a ratio of $3:1$. This undersampling is used to address the notable asymmetry in the number of fraudulent and genuine transactions in the original dataset. The test dataset is then completed by adding all the remaining genuine transactions.

\paragraph*{Hybrid network architecture.}
The first section of the network is composed of a series of classical fully connected feedforward layers. Here, an input layer accepts the first $10$ features. This is followed by two hidden layers of the same size and the result is output on a layer of size $14$. An exponential linear unit (ELU) was used as the nonlinearity.
The second section of our architecture is a quantum neural network consisting of two modes initially in the vacuum. An input layer first operates on the two modes. The input layer omits the first interferometer as this has no effect on the vacuum qumodes. This results in the layer being described by $14$ free parameters, which are set to be directly controlled by the output layer of the classical neural network. The input layer then feeds onto four hidden layers with fully controllable parameters, followed by an output layer in the form of a photon number measurement. An output encoding is fixed in the Fock basis by post-selecting on single-photon outputs and associating a photon in the first mode with a genuine transaction and a photon in the second mode with a fraudulent transaction.

\paragraph*{Training.}
To train the hybrid network, we perform SGD with a batch size of 24. 
Let $p$ be the probability that a single photon is observed 
in the mode corresponding to the correct label for the input transaction. The cost function to minimize is
\beq
C = \sum_{i\in\text{data}} (1-p_i)^2,
\eeq
where $p_i$ is the probability of the single photon being detected in the correct mode on input $i$. The probability included in the cost function is not post-selected on single photon outputs, meaning that training learns to output a useful classification as often as possible. 
We perform training with a cutoff dimension of 10 in each mode for approximately $5 \times 10^{4}$ batches. 
Once trained, we use the probabilities post-selected on single photon events as classification, which could be estimated experimentally by averaging the number of single-photon events occurring across a sequence of runs.

\begin{center}
\begin{figure*}[t!]
\begin{minipage}{0.49\textwidth}
\includegraphics[width=1.1 \columnwidth]{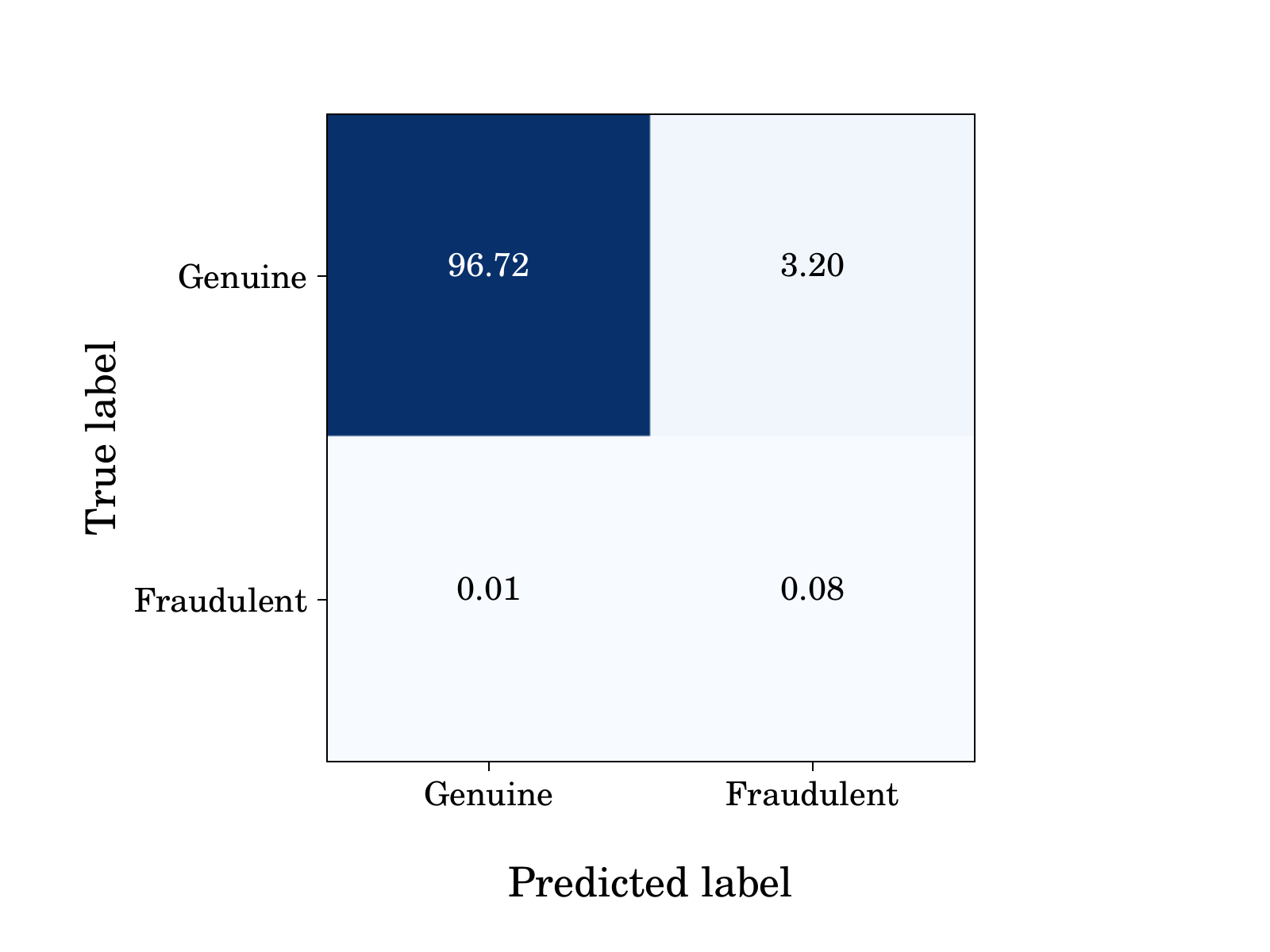}
\end{minipage}
\begin{minipage}{0.49\textwidth}
\vspace{-0.38cm}
\includegraphics[width=1.02 \columnwidth]{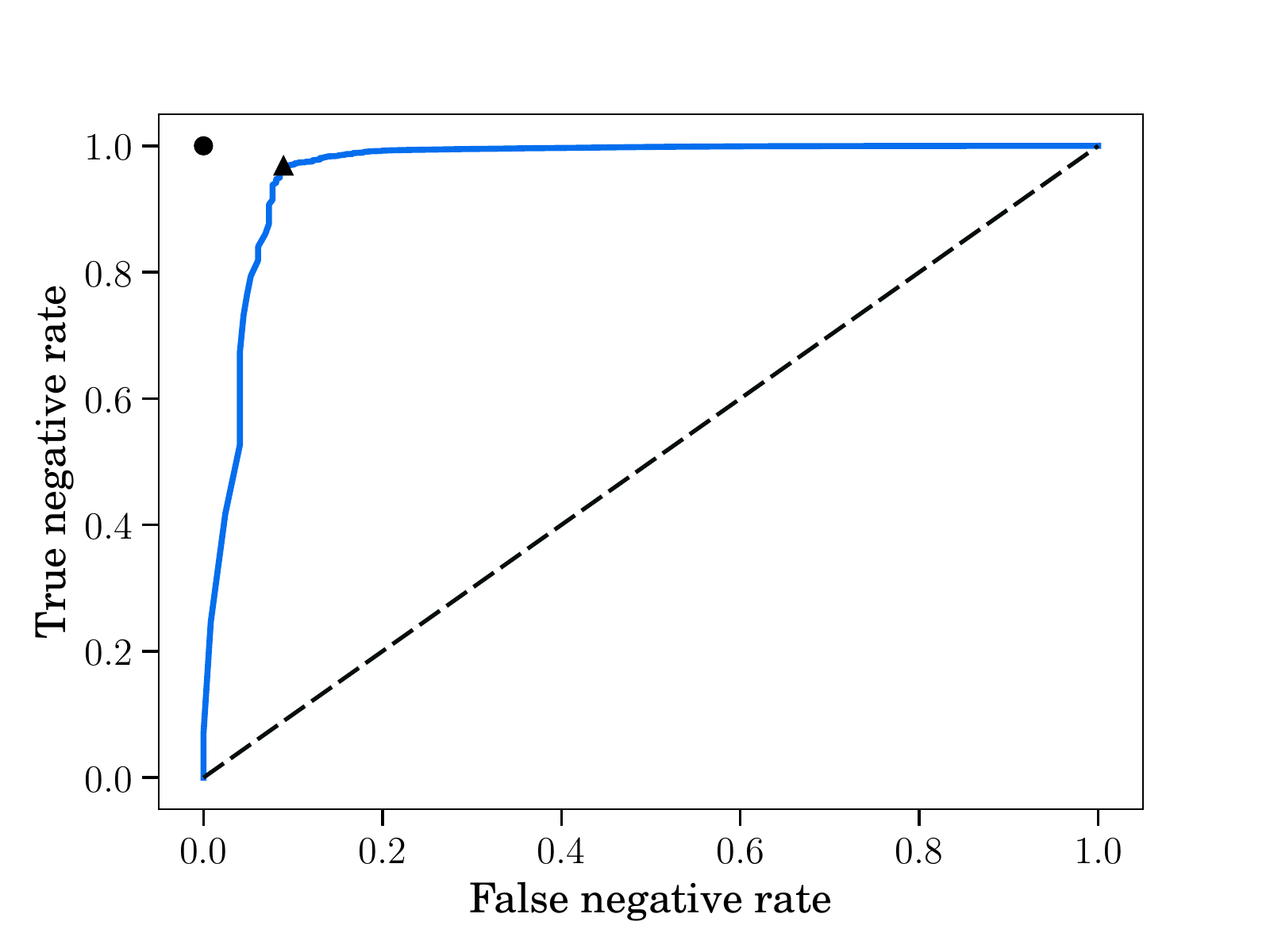}
\end{minipage}
\caption{Experiment B. (Left) Confusion matrix for the test dataset with a threshold probability of $p_{\text{th}}=0.61$. (Right) Receiver operating characteristic (ROC) curve for the test dataset, showing the true negative rate against the false negative rate as a parametric plot of the threshold probability. 
Here, the ideal point is given by the circle in the top-left corner, while the triangle denotes the closest point to optimal among chosen thresholds. This point corresponds to the confusion matrix given here, with threshold $p_{\text{th}}=0.61$.}\label{Fig:FraudResults}
\end{figure*}
\end{center}

\paragraph*{Model performance.}
We test the model by choosing a threshold probability required for transactions to be classified as genuine. The confusion matrix for a threshold of $p_{\text{th}}=0.61$ is given in Fig.~\ref{Fig:FraudResults}. By varying the classification threshold, a receiver operating characteristic (ROC) curve can be constructed, where each point in the curve is parametrized by a value of the threshold. This is shown in Fig.~\ref{Fig:FraudResults}, where the true negative rate is plotted against the false negative rate. An ideal classifier has a true negative rate of $1$ and a false negative rate of $0$, as illustrated by the circle in the figure. Conversely, randomly guessing at a given threshold probability results in the dashed line in the figure. 
Our classifier has an area under the ROC curve of $0.963$, compared to the optimal value of $1$.

For detection of fraudulent credit card transactions, it is imperative to minimize the false negative rate (bottom left square in the confusion matrix of Fig.~\ref{Fig:FraudResults}), i.e., the rate of
misclassifying a fraudulent transaction as genuine. Conversely, it is less important to minimize the false positive rate (top right square) -- these are the cases of genuine transactions being classed as fraudulent. Such cases can typically be addressed by sending verification messages to cardholders. 
The larger false positive rate in Fig.~\ref{Fig:FraudResults} can also be attributed to the large asymmetry between the number of genuine and fraudulent data points.

The results here illustrate a proof-of-principle hybrid classical-quantum neural network able to perform classification for a problem of genuine practical interest.  While it is simple to construct a classical neural network to outperform this hybrid model, our network is restricted in both width and depth due to the need to simulate the quantum network on a classical device. It would be interesting to further explore the performance of hybrid networks in conjunction with a physical quantum computer.

\subsection{Generating images from labeled data}
\label{sec:tetrominos}

\begin{center}
\begin{figure*}[t!]
\begin{tabular}{ccccccc}
\includegraphics[width=0.28 \columnwidth]{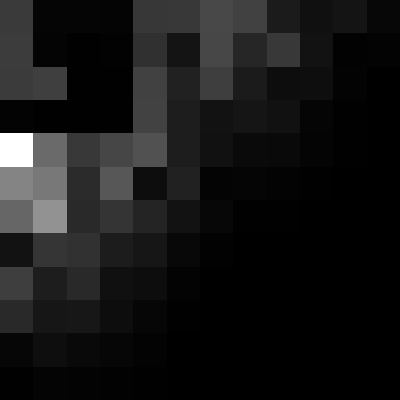}&
\includegraphics[width=0.28 \columnwidth]{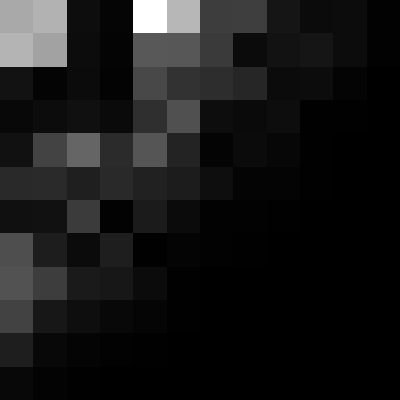}&
\includegraphics[width=0.28 \columnwidth]{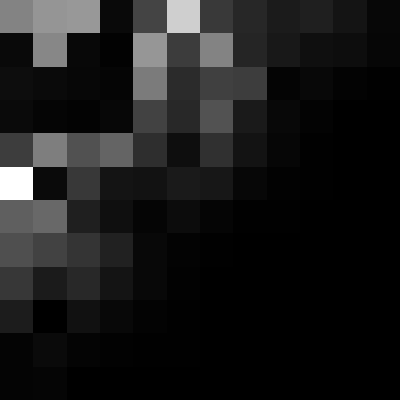}&
\includegraphics[width=0.28 \columnwidth]{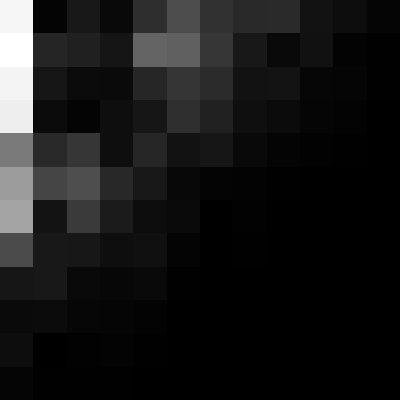}&
\includegraphics[width=0.28 \columnwidth]{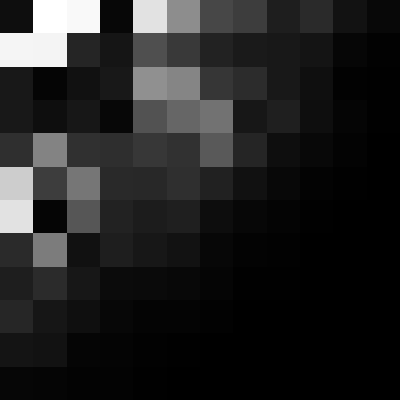}&
\includegraphics[width=0.28 \columnwidth]{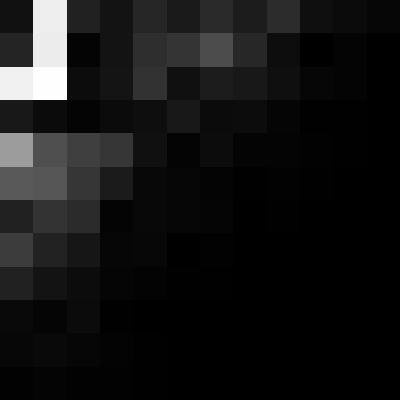}&
\includegraphics[width=0.28 \columnwidth]{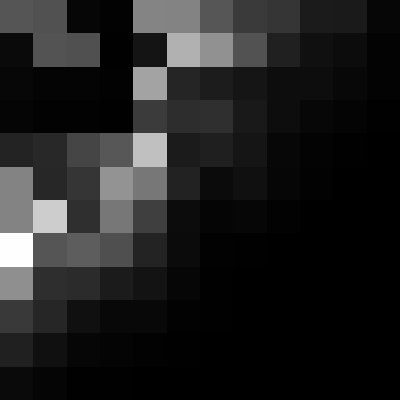}\\
\includegraphics[width=0.28 \columnwidth]{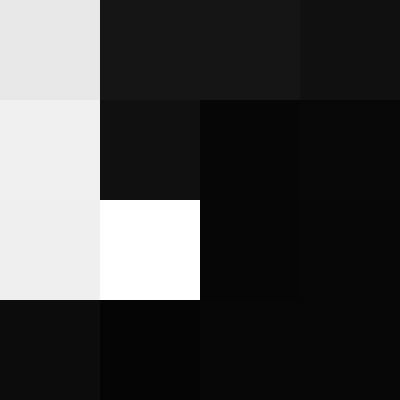}&
\includegraphics[width=0.28 \columnwidth]{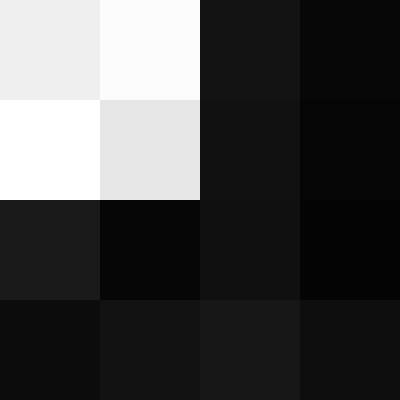}&
\includegraphics[width=0.28 \columnwidth]{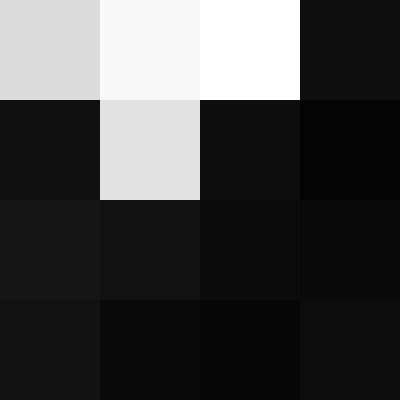}&
\includegraphics[width=0.28 \columnwidth]{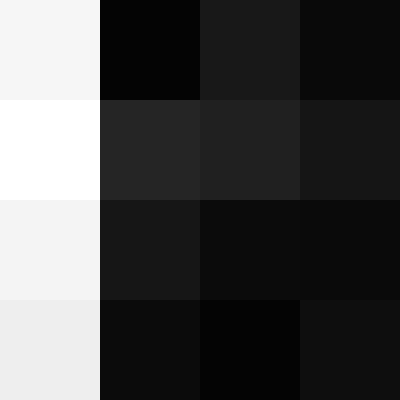}&
\includegraphics[width=0.28 \columnwidth]{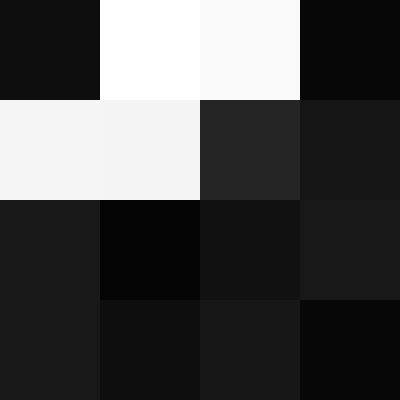}&
\includegraphics[width=0.28 \columnwidth]{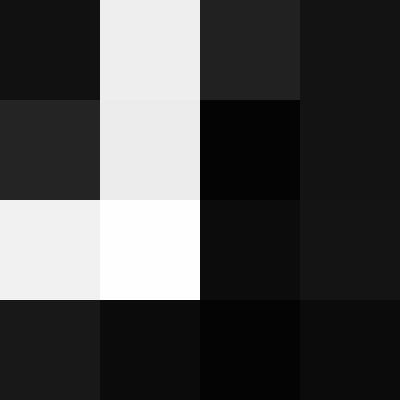}&
\includegraphics[width=0.28 \columnwidth]{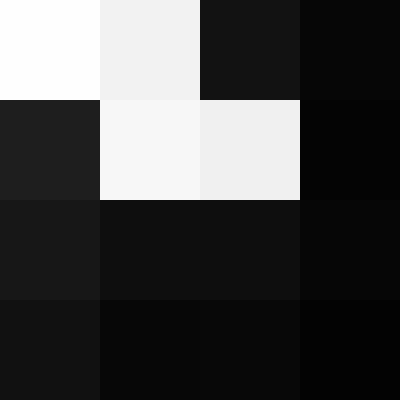}
\end{tabular}
\caption{Experiment C. Ouput images for the `LOTISJZ' tetromino image data. The top row shows the output two-mode states where the intensity of the pixel in the $i$th row and $j$th column is proportional to the probability of finding $i$ photons in the first mode and $j$ photons in the second mode. The bottom row is a close-up in the image Hilbert space of up to 3 photons, renormalized with respect to the probability of projecting the state onto that subspace. In other words, this row illustrates the states $\ket{\psi_i}$ of Eq. \eqref{Eq:psi_outputs}. The fidelities of the output states $\ket{\psi_i}$ with respect to the desired image states are respectively 99.0\%, 98.6\%, 98.6\%, 98.1\%, 98.0\%, 97.8\%, and 98.8\% for an average fidelity of 98.4\%. The probabilities $p_i$ of projecting the state onto the image space of at most three photons are respectively 5.8\%, 36.0\%, 21.7\%, 62.1\%, 40.7\%, 71.3\%, and 5.6\% . }\label{Fig:tetrominos}
\end{figure*}
\end{center} 

Next, we study the problem of training a quantum neural network to generate quantum states that encode grayscale images. We consider images of $N\times N$ pixels specified by a matrix $A$ whose entries $a_{ij}\in [0,1]$ indicate the intensity of the pixel on the $i$th row and $j$th column of the picture. These images can be encoded into two-mode quantum states $\ket{A}$ by associating each entry of the matrix with the coefficients of the state in the Fock basis:
\beq
\ket{A}=\frac{1}{\sqrt{\mathcal{N}}}\sum_{i,j=0}^{N-1} \sqrt{a_{ij}}\ket{i}\ket{j},
\eeq
where $\mathcal{N}=\sum_{i,j=0}^{N-1} |a_{ij}|^2$ is a normalization constant. We refer to these as \emph{image states}. The matrix coefficients $a_{ij}$ are the probability amplitude of observing $i$ photons in the first mode and $j$ photons in the second mode. Therefore, given many copies of a state $\ket{A}$, the image can be statistically reconstructed by averaging photon detection events at the output modes. This architecture is illustrated in Fig. \ref{Fig:Architectures}(c).

\paragraph*{Image encoding strategy.}
Given a collection of images $A_1,A_2, \ldots, A_n$, we fix a set of input two-mode coherent states $\ket{\alpha_1}\ket{\beta_1}, \ket{\alpha_2}\ket{\beta_2}, \ldots, \ket{\alpha_n}\ket{\beta_n}$. The goal is to train the quantum neural network to perform the transformation $\ket{\alpha_i}\ket{\beta_i}\rightarrow \ket{A_i}$ for all $i=1,2,\ldots, n$. Since the transformation is unitary, the Gram matrix of input and output states must be equal, i.e., it must hold that 
\beq
\braket{\alpha_i}{\alpha_j}\braket{\beta_i}{\beta_j}=\braket{A_i}{A_j}
\eeq
for all $i,j$. 

In general, it is not possible to find coherent states that satisfy this condition for arbitrary collections of output states. To address this, we consider output states with support in regions of larger photon number and demand that their projection onto the image Hilbert space of at most $N-1$ photons in each mode coincides, modulo normalization, with the desired output states. Mathematically, if $\mathcal{V}$ is the unitary transformation performed by the quantum neural network, the goal is to train the circuit to produce output states $\mathcal{V}\ket{\alpha_i}\ket{\beta_i}$ such that 
\beq
\Pi_N \,\mathcal{V}\ket{\alpha_i}\ket{\beta_i}=\sqrt{p_i}\ket{A_i},
\eeq
where $\Pi_N = \sum_{i,j=0}^{N-1}\ket{i}\bra{i}\otimes \ket{j}\bra{j}$ is a projector onto the Hilbert space of at most $N-1$ photons in each mode and $p_i=\text{Tr}[\Pi_N \, \mathcal{V}\ket{\alpha_i}\bra{\alpha_i}\otimes\ket{\beta_i}\bra{\beta_i}\mathcal{V}^\dagger]$ is the probability of observing the state in the subspace defined by this projector. The quantum neural network therefore needs to learn not only how to transform input coherent states into image states, it must also learn to employ the additional dimensions in Hilbert space to satisfy the constraints imposed by unitarity. This approach still allows us to retrieve the encoded image by performing photon counting, albeit with a penalty of $p_i$ in the sampling rate.

As an example problem, we select a database of $4\times 4$ images corresponding to the seven standard configurations of four blocks used in the digital game Tetris. These configurations are known as tetrominos. For a fixed value of the parameter $\alpha>0$, the seven input states are set to
\begin{align*}
\ket{\varphi_1}&=\ket{\alpha}\ket{\alpha}\\
\ket{\varphi_2}&=\ket{-\alpha}\ket{-\alpha}\\
\ket{\varphi_3}&=\ket{\alpha}\ket{-\alpha}\\
\ket{\varphi_4}&=\ket{-\alpha}\ket{\alpha}\\
\ket{\varphi_5}&=\ket{i\alpha}\ket{i\alpha}\\
\ket{\varphi_6}&=\ket{-i\alpha}\ket{-i\alpha}\\
\ket{\varphi_7}&=\ket{i\alpha}\ket{\alpha},
\end{align*}
each of which must be mapped to the image state of a corresponding tetromino. 

\begin{figure*}
\begin{minipage}{0.40\textwidth}
\hspace{-2.4cm}
\includegraphics[width=1.0\columnwidth]{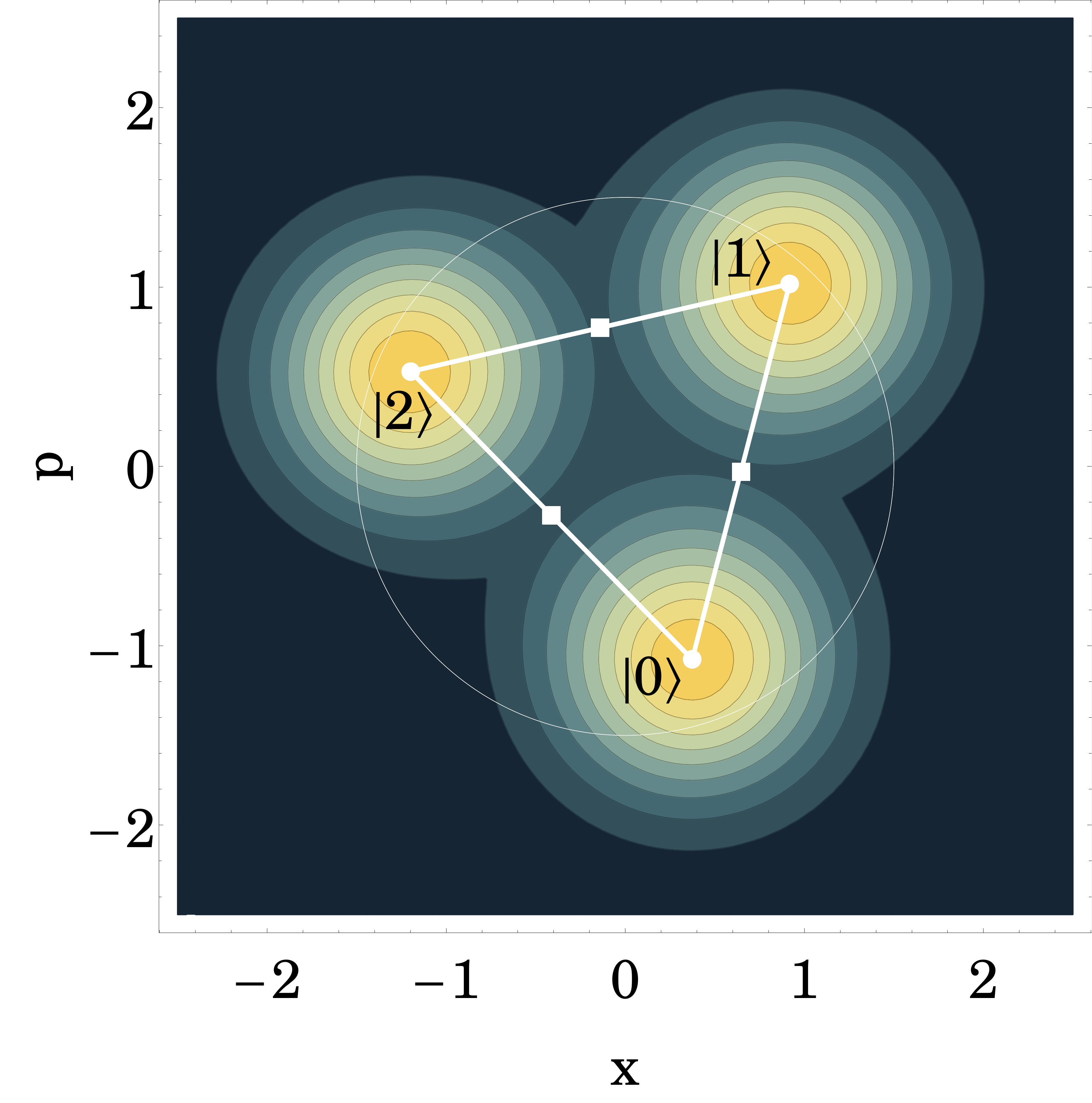}
\end{minipage}
\begin{minipage}{0.028\textwidth}
\hspace{-2.1cm}
\vspace{1.20cm}
\includegraphics[width=1.1\columnwidth]{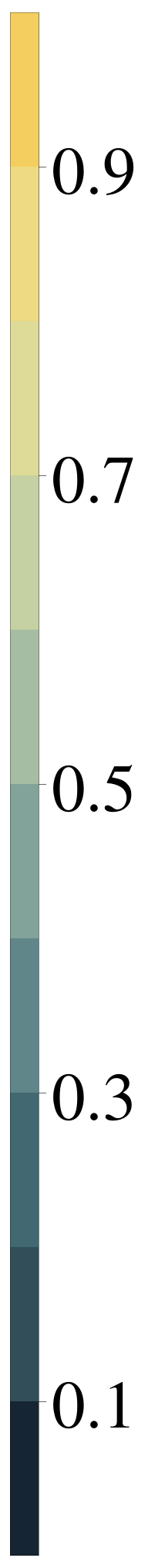}
\end{minipage}
\hspace{-1cm}
\begin{minipage}{0.45\textwidth}
\vspace{-1.2cm}
\includegraphics[width=1.3\columnwidth]{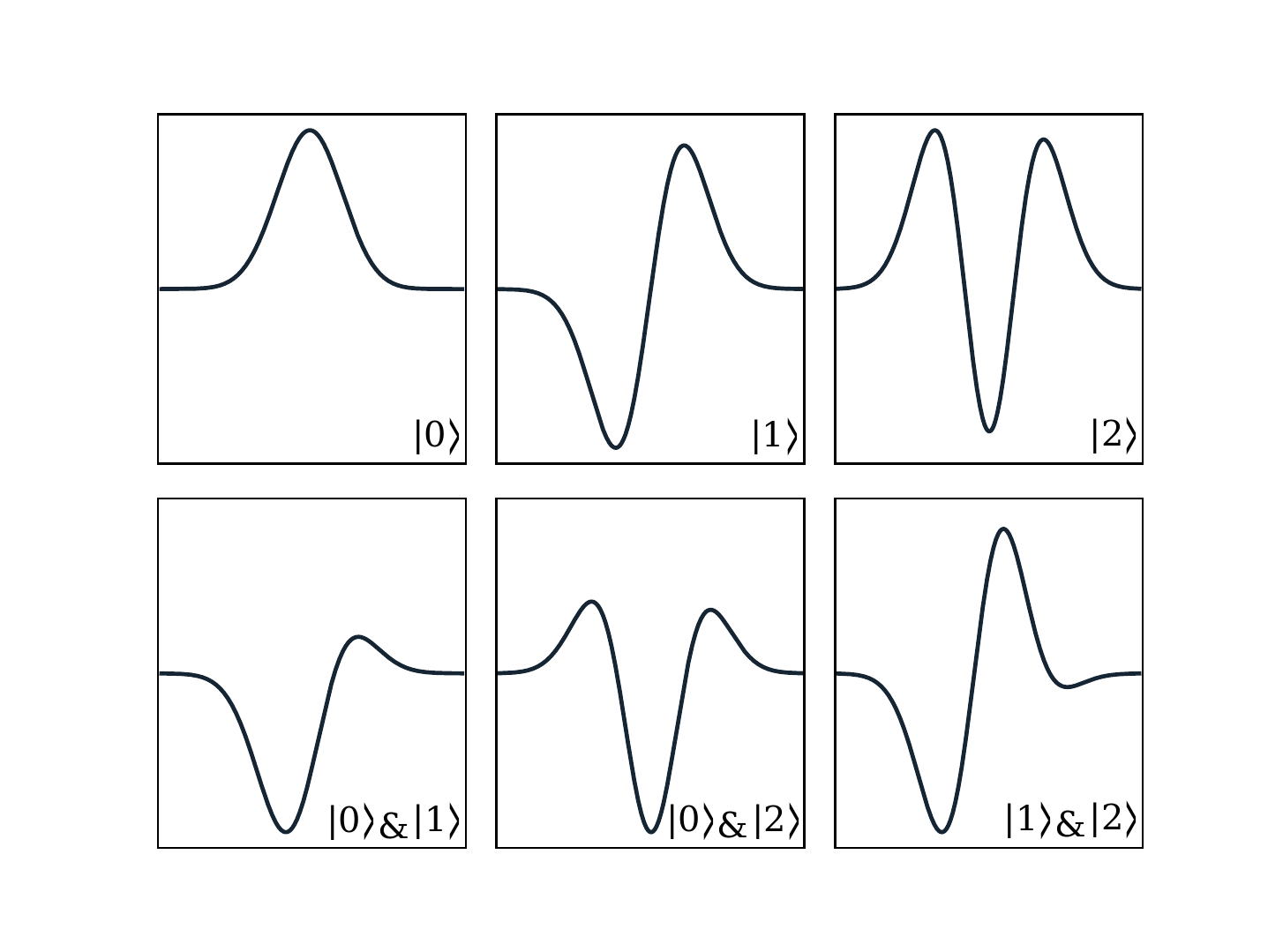}
\end{minipage}
\caption{Experiment D. (Left) Learning a continuous phase-space encoding of the Fock states. The quantum decoder element of a trained classical-quantum autoencoder can be investigated by varying the displacement on the vacuum, which represents the chosen encoding method. The hybrid network has learned to encode the Fock states in different regions of phase space. 
This is illustrated by a contour plot showing, for each point in phase space, the largest fidelity between the output state for that displacement and the first three Fock states.
The thin white circle represents a clipping applied to input displacements during training, i.e., so that no displacement can ever reach outside of the circle. The white circles at points $(0.37, -1.08)$, $(0.92, 1.02)$, and $(-1.20, 0.53)$ represent the input displacements leading to optimal fidelities with the $\ket{0}$, $\ket{1}$, and $\ket{2}$ Fock states, the white lines represent the lines interpolating these optimal displacements, and the white squares represent the halfway points. (Right) Visualizing the wavefunctions of output states. The top row represents the position wavefunctions of states with highest fidelity to $\ket{0}$, $\ket{1}$, and $\ket{2}$, respectively. The bottom row represents the wavefunctions of states with intermediate displacements between the points corresponding to $\ket{0}$ and $\ket{1}$, $\ket{0}$ and $\ket{2}$, $\ket{1}$ and $\ket{2}$, respectively. Each wavefunction is rescaled so that the maximum in absolute value is $\pm 1$, while the $x$ axis denotes positions in the range $[-4.5,4.5]$.
}\label{Fig:FockMusicForever}
\end{figure*}

\paragraph*{Training.}
We define the states
\begin{align}
\ket{\Psi_i}&:=\mathcal{V}\ket{\varphi_i},\\
\ket{\psi_i}&:=\frac{\Pi_4\ket{\Psi_i}}{\|\Pi_4\ket{\Psi_i}\|},\label{Eq:psi_outputs}
\end{align}
i.e., $\ket{\Psi_i}$ is the output state of the network and $\ket{\psi_i}$ is the normalized projection of the output state onto the image Hilbert space of at most 3 photons in each mode. To train the quantum neural network, we define the cost function
\beq
C = \sum_{i=1}^7 |\braket{\psi_i}{A_i}|^2+\gamma P(\{\ket{\Psi_i}\}),
\eeq
where $\ket{A_1},\ket{A_2},\ldots,\ket{A_7}$ are the image states of the seven tetrominos, $P$ is the trace penalty as in Eq. \eqref{Eq: Regularization} and we set $\gamma=100$. By choosing this cost function we are forcing each input to be mapped to a specific image of our choice. In this sense, we can view the images as labeled data of the form $(\ket{\varphi_i}, \ket{A_i})$ where the label specifies which input state they correspond to. 
We employed a network with 25 layers (see Fig.~\ref{Fig:Architectures}(c)) and fixed a cutoff of 11 photons in the numerical simulation, setting the displacement parameter of the input states to $\alpha=1.4$. 

\paragraph*{Model performance.}
The resulting image states are illustrated in Fig. \ref{Fig:tetrominos}, where we plot the absolute value squared of the coefficients in the Fock basis as grayscale pixels in an image. Tetrominos are referred to in terms of the letter of they alphabet they resemble. We fixed the desired output images according to the sequence `LOTISJZ' such that the first input state is mapped to the tetromino `L', the second to `O', and so forth.

Fig.~\ref{Fig:tetrominos} clearly illustrates the role of the higher-dimensional components of the output states in satisfying the constraints imposed by unitarity: the network learns not only how to reproduce the images in the smaller Hilbert space but also how to populate the remaining regions in order to preserve the pairwise overlaps between states. For instance, the input states $\ket{\varphi_1}$ and $\ket{\varphi_2}$ are nearly orthogonal, but the images of the `L' and `O' tetrominos have a significant overlap. Consequently, the network learns to assign a relatively small probability of projecting onto the image space while populating the higher photon sectors in orthogonal subspaces. Overall, the network is successful in reproducing the images in the space of a few photons, precisely as it was intended to do. 

\subsection{Hybrid quantum-classical autoencoder}
\label{sec:fockmusic}
In this example, we build a joint quantum-classical autoencoder (see Fig.~\ref{Fig:Architectures}(d)). Conventional autoencoders are neural networks consisting of an encoder network followed by a decoder network. The objective is to train the network to act as an identity operation on input data. During training, the network learns a restricted encoding of the input data -- which can be found by inspecting the small middle layer which links the encoder and decoder.
For the hybrid autoencoder, our goal is to find a continuous phase-space encoding of the first three Fock states $\ket{0}$, $\ket{1}$, and $\ket{2}$. Each of these states will be encoded into the form of displaced vacuum states, then decoded back to the correct Fock state form.

\paragraph*{Model architecture.} 
For the hybrid autoencoder, we fix a classical feedforward architecture as an encoder and a sequence of layers on one qumode as a decoder, as shown in Fig.~\ref{Fig:Architectures}(d). The classical encoder begins with an input layer with three dimensions, allowing for any real linear combination in the $\{\ket{0}, \ket{1}, \ket{2}\}$ subspace to be input into the network. 
The input layer is followed by six hidden layers of dimension five and a two-dimensional output layer. We use a fully connected model with an ELU nonlinearlity.

The two output units of the classical network are used to set the $x$ and $p$ components of a displacement gate acting on the vacuum in one qumode. This serves as a continuous encoding of the Fock states as displaced vacuum states. In fact, displaced vacuum states have Gaussian distributions in phase space, so the network has a resemblance to a variational autoencoder \cite{kingma2013auto}. We employ a total of $25$ layers with controllable parameters. The goal of the composite autoencoder is to physically generate the Fock state originally input into the network. Once the autoencoder has been trained, by removing the classical encoder we are left with a method to generate Fock states by varying the displacement of the vacuum. Notably, there is no need to specify which displacement should be mapped to each Fock state: this is automatically taken care of by the autoencoder.

\paragraph*{Training.} 
Our hybrid network is trained in the following way. For each of the Fock states $\ket{0}$, $\ket{1}$, and $\ket{2}$, we input the corresponding one-hot vectors $(1, 0, 0)$, $(0, 1, 0)$ and $(0, 0, 1)$ into the classical encoder. Suppose that for an input $\ket{i}$ the encoder outputs the vector $(x_{i}, p_{i})$. This is used to displace the vacuum in one mode, i.e., enacting $\mathcal{D}(\alpha_{i})\ket{0}$ with $\alpha_{i} = (x_{i}, y_{i})$. The output of the quantum decoder is the quantum state $\ket{\Psi_{i}} = \mathcal{V} \mathcal{D}(\alpha_{i})\ket{0}$, with $\mathcal{V}$ the unitary resulting from the layers. We define the normalized projection
\begin{equation}
\ket{\psi_{i}} = \frac{\Pi_{3}\ket{\Psi_{i}}}{\|\Pi_{3}\ket{\Psi_{i}}\|}
\end{equation}
onto the subspace of the first three Fock states, with $\Pi_{3}$ being the corresponding projector. As we have discussed previously, this allows the network to output the state $\ket{\psi_{i}}$ probabilistically upon a successful projection onto the subspace. The objective is to train the network so that $\ket{\psi_{i}}$ is close to $\ket{i}$, where closeness is measured using the fidelity $\left|\langle i|\psi_{i}\rangle\right|^{2}$. As before, we introduce a trace penalty and set a cost function given by
\begin{equation}
C = \sum_{i=0}^{2} \left(\left|\langle i|\psi_{i}\rangle\right|^{2} - 1\right)^{2} +  \gamma P(\{\ket{\Psi_i}\}),
\end{equation}
with $\gamma = 100$ for the regularization parameter. Additionally, we constrain the displacements in the input phase space to a circle of radius $|\alpha|=1.5$ to make sure the encoding is as compact as possible. 

\paragraph*{Model performance.}
After training, the classical encoder element can be removed and we can analyze the quantum decoder by varying the displacements $\alpha$  applied to the vacuum. Fig.~\ref{Fig:FockMusicForever} illustrates the resulting performance by showing the maximum fidelity between the output of the network and each of the three Fock states used for training. For the three Fock states $\ket{0}$, $\ket{1}$, and $\ket{2}$, the best matching input displacements each lead to a decoder output state with fidelity of $99.5 \%$.

The hybrid network has learned to associate different areas of phase space with each of the three Fock states used for training. It is interesting to investigate the resultant output states from the quantum network when the vacuum is displaced to intermediate points between the three areas. These displacements can result in states that exhibit a transition between the Fock states. We use the wavefunction of the output states to visualize this transition. We plot on the right-hand side of Fig.~\ref{Fig:FockMusicForever} the output wavefunctions which give best fidelity to each of the three Fock states $\ket{0}$, $\ket{1}$, $\ket{2}$, respectively. Wavefunctions are also plotted for displacements which are the intermediate points between those corresponding to: $\ket{0}$ and $\ket{1}$; $\ket{0}$ and $\ket{2}$; and $\ket{1}$ and $\ket{2}$, respectively. These plots illustrate a smooth transition between the encoded Fock states in phase space.

\section{Conclusions}
\label{sec:conclusion}

We have presented a quantum neural network architecture which leverages the continuous-variable formalism of quantum computing, and explored it in detail through both theoretical exposition and numerical experiments. This scheme can be considered as an analogue of recent proposals for neural networks encoded using classical light \cite{shen2017deep}, with the additional ingredient that we leverage the quantum properties of the electromagnetic field. Interestingly, as light-based systems are already used in communication networks (both classical and quantum), an optical CV neural network could be wired up directly to communication channels, allowing us to avoid the costly interconversion of classical and quantum information.

We have proposed variants for several well-known classical neural networks, specifically fully connected, convolutional, recurrent, and residual networks. 
We envision that in future work specialized neural networks will also be inspired purely from the quantum side.
We have numerically analyzed the performance of quantum neural network models and demonstrated that they show promise in the tasks we considered. In several of these examples, we employed joint architectures, where classical and quantum networks are used together. This is another promising direction for future exploration, in particular given the current technological lead of classical computers and the expectation that near-term quantum hardware will be limited in size. The quantum part of the model can be specialized to process classically difficult parts of a larger computational to which it is naturally suited. In the longer term, as larger-scale quantum computers are built, the quantum component could take a larger role in hybrid models. Finally, it would be a fruitful research direction to explore the role that fundamental quantum physics concepts -- such as symmetry, interference, entanglement, and the uncertainty principle -- play in quantum neural networks more deeply. 

\acknowledgments We thank Krishna Kumar Sabapathy, Haoyu Qi, Timjan Kalajdzievski, and Josh Izaac for helpful discussions. SL was supported by the ARO under the Blue Sky program.

\appendix
\section{Linear interferometers}
\label{sec:interferometer_proof}
In this section, we derive Eq. (\ref{eq:passive_interferometer_effect}) for the effect of a passive interferometer on the eigenstates $\ket{\bx}$.
A simple expression for an eigenstate of the $\x$ quadrature with eigenvalue $x$ can be found in Appendix 4 of Ref. \cite{barnett2002methods}
\begin{align}
\ket{x} = \pi^{-1/4} \exp\left(-\frac{1}{2} x^2 +\sqrt{2} x \hat a - \frac{1}{2} a^{\dagger 2} \right) \ket{0},
\end{align}
where $\a=\frac{1}{\sqrt{2}}(\x+i\p)$ is the bosonic annihilation operator, and $\ket{0}$ is the single mode vacuum state. The last expression is independent of any prefactors used to define the quadrature operator $\x$ in terms of $\a$ and $\ad$.

This can be easily generalized to $N$ modes:
\begin{align}
	\ket{\bx} 
	& = \bigotimes_{i=1}^N \ket{x_i} \nonumber \\
	& = \pi^{-\frac{N}{4}} \exp\left(
	-\frac{1}{2} \bx^T \bx+\sqrt{2} \bx^T \bad - \frac{1}{2} (\bad)^T \bad
	\right) \ket{\mathbf{0}},
\end{align}
where now
\begin{align}
\bx &= (x_1,\ldots,x_N)^T, \\
\bad &= (\ad_1,\ldots,\ad_N)^T, \quad (\bad)^T = (\ad_1,\ldots,\ad_N),
\end{align}
and $\ket{\mathbf{0}}$ is the multimode vacuum state.
Now consider a (passive) linear optical transformation $\mathcal{U}$
\begin{align}
\label{transformations}
	\ad_i &\to \mathcal{U} \ad_i \mathcal{U}^\dagger = \sum_{j} U_{i j} \ad_j , \\
	\bad &\to U \bad, \quad 	\left(\bad\right)^T\to \left(\bad\right)^T U^T.
\end{align}
In general, $U$ is an arbitrary unitary matrix, $U U^\dagger = \mathbbm{1}_N$. We will however restrict $U$ to have real entries and thus to be orthogonal. In this case, $U^\dagger = U^T$ and hence $U^T U = U U^T = \mathbbm{1}_N$.

We can now examine how the multimode state $\ket{\bx}$ transforms under such a linear interferometer $\mathcal{U}$:
\begin{align}
& \mathcal{U} \ket{\bx}  \\
& = \mathcal{U}  \left[ \pi^{-\frac{N}{4}} \exp\left(
-\tfrac{1}{2} \bx^T \bx+\sqrt{2} \bx^T \bad - \tfrac{1}{2} (\bad)^T \bad
\right) \ket{\mathbf{0}} \right] \nonumber \\
& = \mathcal{U} \frac{   \exp\left(
-\tfrac{1}{2} \bx^T \bx+\sqrt{2} \bx^T \bad - \tfrac{1}{2} (\bad)^T \bad
\right)}{\pi^{\frac{N}{4}} } \mathcal{U}^\dagger \mathcal{U} \ket{\mathbf{0}} \nonumber \\
& = \frac{    \exp\left( \mathcal{U} \left[
-\tfrac{1}{2} \bx^T \bx+\sqrt{2} \bx^T \bad - \tfrac{1}{2} (\bad)^T \bad \right] \mathcal{U}^\dagger
\right) }{\pi^{\frac{N}{4}}}  \ket{\mathbf{0}} . \nonumber
\end{align}
We can use the transformation in Eq. (\ref{transformations}) to write 
\begin{align}
\mathcal{U} \ket{\bx} =&    \frac{\exp\left(
-\tfrac{1}{2} \bx^T \bx+\sqrt{2} \bx^T U \bad - \tfrac{1}{2} (\bad)^T U^T U \bad 
\right) }{\pi^{\frac{N}{4}} }  \ket{\mathbf{0}}.
\end{align}
Now we use that  $U^T U = U U^T=  \mathbbm{1}_N$ to write the last expression as
\begin{align}
\mathcal{U} \ket{\bx} =&  \frac{ \exp\left(
-\frac{1}{2} \bx^T U U^T \bx+\sqrt{2} \bx^T U \bad - \frac{1}{2} (\bad)^T  \bad  
\right)}{\pi^{\frac{N}{4}}  }   \ket{\mathbf{0}} .
\end{align}
Let us define the vector $\bm{y} = U^T \bx$ and, to match the notation of Eq. (\ref{eq:passive_interferometer_effect}), the orthogonal matrix $C=U^T$, in terms of which we find
\begin{align}
\mathcal{U} \ket{\bx} \nonumber \\
=& \frac{   \exp\left(
-\frac{1}{2} \bx^T U U^T \bx+\sqrt{2} \bx^T U \bad - \frac{1}{2} (\bad)^T  \bad
\right) }{\pi^{\frac{N}{4}} }  \ket{\mathbf{0}} \nonumber \\
=& \frac{   \exp\left(  
-\frac{1}{2} \bm{y}^T \bm{y}+\sqrt{2} \bm{y}^T \bad - \frac{1}{2} (\bad)^T  \bad
\right) }{\pi^{\frac{N}{4}} }  \ket{\mathbf{0}} \nonumber \\
=& \ket{\bm{y}} \nonumber \\
=& \ket{C \bx}.
\end{align}

Note that the output state is also a product state. This simple product transformation is a corollary of the elegant results of Ref. \cite{jiang2013mixing}: ``Given a nonclassical pure-product-state input to an $N$-port linear-optical network, the output is almost always mode entangled; the only exception is a product of squeezed states, all with the same squeezing strength, input to a network that does not mix the squeezed and antisqueezed quadratures.'' In our context the $x$ eigenstates are nothing but infinitely squeezed states and the fact that our passive linear optical transformation is orthogonal immediately implies that squeezed and antisqueezed quadratures are not mixed.

\section{Convolutional networks}
\label{app:convolutional_proof}
In this section, we derive the connection between a translationally-invariant Hamiltonian and a Block Toeplitz symplectic transformation. The notion of translation symmetry and Toeplitz structure are both connected to one-dimensional convolutions. 
Two-dimensional convolutions, naturally appearing in image processing applications, are connected not with Toeplitz matrices, but with doubly block circulant matrices \cite{goodfellow2016deep}. We will not consider this extension here, but the basic ideas are the same.

Suppose we have a Hamiltonian operator $H=H(\mathbf{\x}, \mathbf{\p})$ which generates a Gaussian unitary $U=\exp(-itH)$ on $N$ modes. We are interested only in the matrix multiplication part of an affine transformation, i.e., $H$ does not generate displacements. Under these conditions, $H$ has to be quadratic in the operators $(\mathbf{\x}, \mathbf{\p})$, 
\begin{equation}
 H = 
 \begin{bmatrix}
  \mathbf{\x}^T & \mathbf{\p}^T
 \end{bmatrix}
 \begin{bmatrix}
  H_{\bx\bx} & H_{\bx\bp} \\
  H_{\bp\bx} & H_{\bp\bp}
 \end{bmatrix}
 \begin{bmatrix}
  \mathbf{\x} \\ 
  \mathbf{\p}
 \end{bmatrix},
\end{equation}
where each $H_{\mathbf{u}\mathbf{v}}$ is an $N\times N$ matrix.
We will call the inner matrix in this equation $\widetilde{H}$.
In the phase space picture, the symplectic transformation $M_H$ generated by $H$ is obtained via the rule \cite{serafini2017quantum}
\begin{equation}
 \label{eq:symplectic_from_hamiltonian}
 M_H = \exp(\Omega \widetilde{H}),
\end{equation}
where $\Omega$ is the symplectic form from Eq. (\ref{eq:symplectic_form}).

We now fix $H$ to be translationally invariant, i.e., $H$ does not change under the transformation 
\begin{equation}
 \begin{bmatrix}
  \mathbf{\x} \\
  \mathbf{\p}
 \end{bmatrix}
 \mapsto
 \begin{bmatrix}
  T\mathbf{\x} \\
  T\mathbf{\p}
 \end{bmatrix},
\end{equation}
where we have introduced the shift operator $T$ which maps $\x_i\mapsto\x_{i+1}$ and $\p_i\mapsto\p_{i+1}$. We assume periodic boundary conditions on the modes, $\x_{N}\mapsto\x_1$ and $\p_{N}\mapsto\p_1$, which allows us to represent translation as an $N\times N$ orthogonal matrix:
\begin{equation}
 T = \sum_i \ketbra{i+1}{i}.
\end{equation}
The translationally-invariant condition on $H$ translates to the statement that
\begin{equation}
 [T, H_{\mathbf{u}\mathbf{v}}] = 0
\end{equation}
for $\mathbf{u},\mathbf{v}\in\{\bx,\bp\}$.

In the $2N$-dimensional phase space, the $N$-dimensional translation matrix takes the form $T\oplus T$. Considering the expression
\begin{equation}
 [\Omega\widetilde{H},T\oplus T ] = 
 \begin{bmatrix}
  [M_{\bp\bx}, T] & [M_{\bp\bp}, T] \\
  -[M_{\bx\bx}, T] & -[M_{\bx\bp}, T]
 \end{bmatrix}
 = 0,
\end{equation}
we see that the symplectic matrix $M_H$ from Eq. (\ref{eq:symplectic_from_hamiltonian}) must also be symmetric under translations:
\begin{equation}
 [M_H, T\oplus T] = 0.
\end{equation}
Writing this matrix in a block form, 
\begin{equation}
 M_H =  
   \begin{bmatrix}
     M_{\bx\bx} & M_{\bx\bp} \\
     M_{\bp\bx} & M_{\bp\bp}
   \end{bmatrix},
\end{equation}
we conclude that we must also have 
\begin{equation}
 [M_{\mathbf{u}\mathbf{v}}, T] = 0
\end{equation}
for each $\mathbf{u},\mathbf{v}\in\{\bx,\bp\}$.
Expressing this in the equivalent form
\begin{equation}
 T^T M_{\mathbf{u}\mathbf{v}} T = M_{\mathbf{u}\mathbf{v}},
\end{equation}
we see that the following condition must hold on the entries of each $M_{\mathbf{u}\mathbf{v}}$:
\begin{equation}
 [M_{\mathbf{u}\mathbf{v}}]_{ij} = [M_{\mathbf{u}\mathbf{v}}]_{i+1,j+1}.
\end{equation}
In other words, when the generating Hamiltonian is translationally invariant, each block of the corresponding symplectic matrix is a Toeplitz matrix, which implements a one-dimensional convolution.

\bibliographystyle{unsrt}
\bibliography{references}

\end{document}